\documentclass[twocolumn,superscriptaddress,nopacs,preprintnumbers,amsmath,amssymb,nofootinbib]{revtex4}

\usepackage{verbatim}
\usepackage{amsmath,amssymb}
\usepackage{graphicx,psfrag}
\newlength{\scalegraph}
\usepackage{bbm}

\def\Sw{S}
\def\Sc{S_{\rm cl}}
\def\Norm{{\cal N}}

\def\s0#1#2{\mbox{\small{$ \frac{#1}{#2} $}}}
\def\0#1#2{\frac{#1}{#2}}

\def\eq#1{(\ref{#1})}
\def\Eq#1{eq.~(\ref{#1})}

\def\ep{\epsilon}
\def\de{\delta}

\def\n{\nabla}
\def\tn{\hat{\nabla}}

\def\p{\phi}
\def\vp{\varphi}
\def\bp{\bar{\psi}}

\def\Sdet{{\rm SDet}}
\def\Str{{\rm STr}}
\def\Di{{\cal D}}
\def\M{M}
\newcommand{\Mc}{\tilde{M}}

\newcommand{\dec}{\tilde{\delta}}

\newcommand{\defgamma}{\gamma_{5,{\rm def}}}
\newcommand{\bardefgamma}{\bar\gamma_{5,{\rm def}}}

\newcommand{\half}{\frac{1}{2}}
\newcommand{\bpsi}{\bar{\psi}}
\newcommand{\pr}{\prime}

\newcommand{\alm}{\alpha^{-1}}

\newcommand{\al}{\alpha}

\newcommand{\sy}[1]{#1}
\newcommand{\asy}[1]{#1''}

\newcommand{\elms}{d_1}
\newcommand{\elm}{b}

\def\d{{\rm d}}
\def\pa{\partial}

\def\Nt{N}

\def\d{{d}}
\def\tr{{\rm tr}}
\def\Tr{{\rm Tr}}

\def\id{1\!\mbox{l}}

\begin{document}

\title{Generalising the Ginsparg-Wilson relation:\\ Lattice
Supersymmetry from Blocking Transformations}

\author{Georg Bergner}
\affiliation{Theoretisch-Physikalisches Institut, Universit\"at Jena,
  D-07743 Jena, Germany} \author{Falk Bruckmann}
\affiliation{Institut f\"ur Theoretische Physik, Universit\"at
  Regensburg, D-93040 Regensburg, Germany} \author{Jan M.~Pawlowski}
\affiliation{Institut f\"ur Theoretische Physik, Universit\"at
  Heidelberg, D-69120 Heidelberg, Germany}

\begin{abstract}
  The Ginsparg-Wilson relation is extended to interacting field
  theories with general linear symmetries.  Our relation encodes the
  remnant of the original symmetry in terms of the blocked fields and
  guides the construction of invariant lattice actions.  We apply this
  approach in the case of lattice supersymmetry.  An additional
  constraint has to be satisfied because of the appearance of a
  derivative operator in the symmetry transformations.  The solution
  of this constraint leads to non-local SLAC-type derivatives. We
  investigate the corresponding kinetic operators on the lattice
  within an exact solution of supersymmetric quantum mechanics.  These
  solutions -- analogues of the overlap operator for supersymmetry --
  can be made local through a specific choice of the blocking kernel.
  We show that the symmetry relation allows for local lattice
  symmetry operators as well as local lattice actions. We argue that
  for interacting theories the lattice action is polynomial in the
  fields only under special circumstances, which is exemplified within
  an exact solution.
\end{abstract}

\maketitle

\section{Introduction}

Lattice simulations of the path integral are a powerful tool to study
quantum field theories, especially their non-perturbative properties.
The first step in this program is to find a correct discretisation of
the continuum action and its symmetries (and appropriate discrete
observables).  In that respect a lot of experience has been collected
in the case of gauge and chiral theories.

There have also been efforts to simulate supersymmetric (SUSY)
theories on the lattice for many years.  For a quadratic, i.e.\ free
theory, a supersymmetric lattice action can be constructed.  However,
one encounters great difficulties in finding more general lattice
actions that are invariant under the (naively) discretised
supersymmetry transformations.  One of them can be traced back to the
fact that SUSY transformations contain derivatives of fields and that
the continuum SUSY actions are invariant up to total derivative terms.
In general the corresponding terms on the lattice do not vanish
because the Leibniz and chain rule of differentiation is violated by
any lattice derivative operator \cite{Dondi:1976tx}.  The optimal
choice in this respect is the SLAC derivative \cite{SLAC}, which
preserves the Leibniz rule in the first Brillouin zone (BZ) (see
\cite{Bergner:2007pu} for recent works), but leads to non-localities
\cite{Karsten:1980wd}.

Simulations without a realisation of supersymmetry on the lattice
\cite{brute} generically suffer from fine-tuning problems in the
continuum limit.  In higher SUSY theories it is possible to realise a
part of the supersymmetry, and one can hope that this partial
realisation already ensures the correct continuum limit. Many
approaches to lattice supersymmetry rely on such constructions, as
e.g.\ \cite{Nilpot}.  Another possibility is to use a (nonlinear)
deformation of the continuum transformations on the lattice
\cite{Feo}.  One then still has to show that such lattice
transformations resemble the SUSY transformations in the continuum
limit even in the presence of quantum corrections to all orders of
perturbation theory, as done for a specific model in~\cite{Golterman}.
This can ensure the correct continuum limit of these deformed
transformations, but a non-perturbative argument would be desirable.
A further attempt to keep the full SUSY by representing it on
non-commutative objects or link objects \cite{D'Adda:2004jb} has been
criticised in \cite{Bruckmann:2006ub}.  For more details on lattice
supersymmetry we refer the reader to \cite{reviews}.  The conclusion
of all these studies seems to be that there is a fundamental obstacle
against interacting local lattice theories with exact SUSY implemented
in the naive way.

This situation resembles very much that of chiral theories on the
lattice.  The Nielsen-Ninomiya theorem \cite{Nielsen} forbids exact
chiral symmetry on the lattice under a few very reasonable assumptions
like locality.  Ginsparg and Wilson have derived a modified symmetry
relation for a chiral lattice theory \cite{Ginsparg:1981bj}; in
particular, the lattice Dirac operator anticommutes with $\gamma_5$
except for a term that vanishes in the continuum limit $a\to 0$.  A
solution to this relation was found by Neuberger
\cite{Neuberger:1998wv}, which is not ultra-local \cite{UltraLocal},
but local if the background gauge field is smooth (the plaquette is
close to the identity) \cite{Hernandez:1998et}.

The Ginsparg-Wilson (GW) relation is not a mere modification of the
naive lattice symmetry with terms that formally vanish in the limit
$a\to 0$; its justification is the analysis of the Wilsonian
renormalisation group \cite{Igarashi:2002ba}, for the general setting
see also
\cite{Pawlowski:2005xe,Litim:1998nf,Bonini:1993kt,Igarashi:2001mf}.
Thus it is a non-perturbative construction of a deformed symmetry
transformation.  Our strategy in putting supersymmetry on the lattice
is therefore to revisit the corresponding procedure (this was already
suggested in \cite{Luscher:1998pqa} and some earlier but incomplete
attempts can be found in \cite{So:1998ya}).  It leads to a lattice
theory in an effective Wilsonian sense.  This can be viewed as
integrating out quantum fluctuations up to a finite number of lattice
degrees of freedom, the integration over which is performed
numerically.  It is well-known that in such a process of quantisation
classical symmetries get deformed.  The aim of the present work is not
only to extend the Ginsparg-Wilson approach to supersymmetry.  We will
also generalise the investigations of Ginsparg and Wilson for an
arbitrary linear symmetry. The reduction of the degrees of freedom is
done with an appropriate blocking transformation and the results will
depend on the choice of a blocking kernel.

If only a quadratic theory is considered the effective lattice action
for a given continuum action can be calculated.  This was done in
\cite{Bietenholz} for a supersymmetric model.  Here we are not
attempting to find such explicit solutions, since this is possible
only for a free theory.  Rather we investigate the implications of a
continuum symmetry for the effective lattice action.  If these
implications are fulfilled a symmetric theory should be approached in
the continuum limit.

As our main result, we derive an exact relation that a lattice theory
has to obey in order to represent the continuum symmetry.  Also
interacting theories are included in this generalisation of the GW
relation to a general linear symmetry.  By contrast, the GW relation
deals only with the chiral symmetry acting on the quadratic fermion
part of the action; the gauge fields are mere spectators.  As in the
GW case the naive lattice symmetry will be modified by terms that are
proportional to the inverse blocking kernel.  If this relation should
represent a proper lattice symmetry certain conditions, especially
locality, must be satisfied. These conditions exclude e.g.\ the
Wilson operator as a solution for chiral symmetry.

Applying the formalism to supersymmetry with its derivative
transformations, an additional constraint has to be fulfilled in order
to derive a lattice remnant of the continuum symmetry.  The solution
of this constraint generically leads to the non-local SLAC operator in
the lattice transformations.  We also investigate the continuum limit
of the blocking kernel, which is less restrictive.

We are able to solve the relation for the quadratic (kinetic plus
mass) sector of supersymmetric quantum mechanics (SUSYQM).  In the
case of this one-dimensional toy model the relation already yields
non-trivial difference operators -- analogues of the overlap operator
for the case of supersymmetry.

The locality of these derivative operators can be improved using the
freedom in the blocking kernel.
This is an important result of our strategy: The blocking kernel helps
in achieving the desired properties of the lattice theory, `at the
expense of' introducing a rhs in the lattice symmetry relation
rendering it different from the naive one.

Concerning interacting SUSY theories, the symmetry relation
generically couples different powers of the fields in the action
beyond second order.  It is therefore very intricate to truncate the
interaction in the power of fields.  As an example displaying these
difficulties we solve the case of constant fields in SUSYQM.  For
general theories we give a necessary criterion for
the construction of polynomial interactions.\\
 
The paper is organised as follows: First we introduce the blocking
procedure in detail and derive the symmetry relation.  It is
particularly simple when a quadratic action is considered, and for
chiral symmetry we recover the GW relation.  The next section is
devoted to the additional constraint and its solutions.  Their
continuum limit as well as the continuum limit of the blocking kernel
is analysed carefully in Sect. \ref{sec:contlim}.

In Sect. \ref{sec:quadact} we start to apply the approach to SUSYQM,
giving the general solution for the quadratic case and discussing its
properties, especially locality.  Afterwards in Sect.
\ref{sec:interacting} we solve the constant field case and investigate
the possibility of polynomial actions fulfilling the symmetry
relation, both in general and for SUSY.  We end with a summary of our
results and a discussion of their implications.

\section{Wilsonian effective action}
\label{sec:Wilson}
In this chapter the Wilsonian effective action computed from a general
blocking transformation is introduced. The key result of the following
investigation are the blocked symmetry transformations. These are
specified later for the case of supersymmetry.  The simple form of the
symmetry transformations in case of a quadratic action is derived in
section~\ref{sec:quadAction}.  With this result the general concepts
are elucidated at the example of chiral symmetry where the well-known
Ginsparg-Wilson relation is reproduced.

\subsection{Blocking}
\label{sec:Wilson_blocking}

The starting point is the generating functional of a general quantum
theory
\begin{eqnarray}\label{eq:genfunc1}
  Z[j]=\0{1}\Norm\int d\varphi \, e^{-\Sc[\varphi]+
    \int j\varphi} 
  \,, 
\end{eqnarray} 
with classical action $\Sc[\varphi]$ and fields
$\varphi(x)=(\varphi^1(x)\,,\,...\,\varphi^A(x))$ comprising both,
bosonic and fermionic degrees of freedom.

The theory is reduced to a finite number of degrees of freedom by
introducing averaged or blocked fields $\phi_f(an)$ at lattices sites
$n\in N_1\times N_2\times\ldots\times N_d$ by a linear blocking
procedure,
\begin{eqnarray}\label{eq:average} 
  \phi_f(an):=\int d^d x\, f(an-x)\, \varphi(x)\,, 
\end{eqnarray} 
where $a$ is the lattice distance, and a blocking function $f$ peaked
at $0$.  $f$ should have the dimension inverse to the
$d$-dimensional integral, such that the original and blocked fields
have the same dimension.

This blocking can be easily generalised to finite blocking steps and
fields $\varphi^i$ where $i$ comprises internal indices and species of
fields. Then a general blocking reads
\begin{eqnarray}\label{eq:blocking1}
  (\phi_{f})^i_n =f^{ij}_{nx} \varphi^j_x\,, 
\end{eqnarray} 
where a summation/integration over indices is understood. The blocking
matrix $f^{ij}_{nx}$ is rectangular, and usually relates only the same
species, but in general it also mixes the internal indices. In
\eq{eq:blocking1} $x$ can be a discrete index, then $\varphi_x$ is
already understood as a blocked continuum field
$\varphi_x=(\phi_{f_0})_x$.  In that case the blocking maps a lattice
with a smaller lattice spacing and consequently more degrees of
freedom onto a smaller one with a larger lattice spacing.

The blocking procedure amounts to rewriting the generating functional
$Z$ in eqn.\ \eq{eq:genfunc1} in terms of a path integral on the
lattice defined by the image of $f$. For the sake of simplicity we
concentrate on the generating functional at vanishing sources for now,
the general case, important for the discussion of the continuum limit,
will be discussed in section~\ref{sec:contlim}. We proceed with
rewriting the generating functional $Z[0]$ as
  \begin{eqnarray} \label{eq:Zphi} 
 Z[0] = \0{1}{\Norm}\int d\phi\,
    e^{-\Sw[\phi]}\, ,
\end{eqnarray} 
where the $\phi_n^i$ live on the lattice given by the blocking $f$,
and $\Sw$ is the Wilsonian effective action with
\begin{eqnarray} \label{eq:SWilson} 
e^{-\Sw[\phi]}=\Sdet{}^{-1/2}\alpha\, \int d\varphi\,
  e^{-\012(\phi-\phi_f) \alpha (\phi-\phi_f)-\Sc[\varphi]}\,.
\end{eqnarray} 
$\Sdet$ is the super-determinant, i.e.\ the determinant for bosons and
its inverse for fermions\footnote{Strictly speaking $\Sdet^{-1/2}\al$
  in our case means the Pfaffian of $\al$ for fermionic fields and
  $\det^{-1/2}\al$ for the bosonic part.}.  In \eq{eq:SWilson} we have
introduced a quadratic smearing with a blocking kernel
\begin{eqnarray}\label{eq:alphascalar} 
(\phi-\phi_f)\alpha(\phi-\phi_f) &=&(\phi-\phi_f)_n^i \alpha^{ij}_{nm}
    (\phi-\phi_f)^i_m \,.
\end{eqnarray}
By inserting \eq{eq:SWilson} into \eq{eq:Zphi} and performing the
Gau\ss ian integration over $\phi$ it can be straightforwardly checked
that \eq{eq:Zphi} gives $Z[0]$ in \eq{eq:genfunc1}. Note that the
above smearing also encompasses $\alpha$'s with diverging or vanishing
determinants. 
This and further details will be
discussed in section~\ref{sec:contlim}. The explicit examples of
quadratic actions and chiral symmetry are shown in
section~\ref{sec:quadAction} and section~\ref{sec:chiral} respectively.

  For illustration of the smearing procedure we briefly discuss a
  specifically simple case with $\alpha_{nm}^{ij}=\alpha\,
  \delta^{ij}\delta_{mn}$ with $\alpha\to \infty$.  Then the smearing
  term turns into a $\delta$-function in field space,
\begin{eqnarray}\label{eq:deltaphi}
\Sdet{}^{-1/2}\alpha\, e^{-\012(\phi-\phi_f)\alpha (\phi-\phi_f)}=
\delta(\phi-\phi_f)\,, 
\end{eqnarray}
and the smearing is removed. Removing the smearing in the continuum
limit, $a\to 0$, is necessary in order to recover the original action
in this limit, $\Sw\to \Sc$, apart from the blocking $f$ becoming the
delta-distribution, that is $\phi_f\to\varphi$.  Furthermore, for
achieving \eq{eq:deltaphi}, $\alpha/a^d$ has to diverge. More general
restrictions for the continuum limit will be addressed in detail in
section~\ref{sec:contlim}.

The blocking kernel shall connect bosons and fermions only among
themselves and in these subspaces it obeys
\begin{eqnarray}
  \alpha^{ij}_{nm}=\alpha^{ji}_{mn} (-1)^{|\p^i||\p^j|}\,,
\label{eq_alpha_symm}
\end{eqnarray} where
\begin{eqnarray} |\p^i|= \left\{\begin{array}{ll} 1 \quad & \phi^i\
{\rm fermionic} \\ 0 \quad & \phi^i \ {\rm bosonic}\end{array}\right.
\label{fermbos}\, .
\end{eqnarray} 
In other words, $\alpha$ is antisymmetric for fermions and symmetric
for bosons, $\alpha=\pm\alpha^T$, where the minus sign applies
whenever fermionic indices are interchanged in the transposition 
of the matrix.\\

\subsection{Symmetries}
\label{sec:derivation}

The primary concern of this construction is the blocking
transformation of symmetries of the classical action.  We now
investigate in what form the lattice action inherits this symmetry.
Our main result will be the relation \eq{eq:SymWilson} that
corresponds to the Ginsparg-Wilson relation, but is valid for a
general linear symmetry.  Continuum implementations of the related
ideas close to the present line of arguments can be found in e.g.\
\cite{Bonini:1993kt,Igarashi:2001mf}, for reviews see
\cite{Pawlowski:2005xe,Litim:1998nf}.

Let the classical action $S$ be invariant under a linear
transformation
\begin{equation}
 \varphi\to\varphi+\dec\varphi\,,\qquad 
 (\dec\varphi)^i_{x} = \ep \Mc_{xy}^{ij}\varphi_{y}^j\,,
\label{eqq:classvar}
\end{equation}
where $\Mc$ in general relates different field species $(i,j)$, but
may also act non-trivially on the coordinates ($x$,$y$).  $\ep$ is the
small parameter of the transformation.  In the application to SUSY,
$\ep$ is Grassmann-valued as $\Mc$ mixes bosons and fermions, and
$\Mc$ also contains derivatives.  We have introduced the notation that
for a given lattice quantity ${\cal O}$ the $\tilde{\cal O}$ refers to
the corresponding continuum quantity.

In combination with the averaging function $f$, see eqn.s
\eq{eq:average} and \eq{eq:blocking1}, this symmetry transformation
induces a corresponding transformation on the blocked field $\phi_f$
\begin{eqnarray}
(\dec \phi_f)^i_n= \epsilon f^{ij}_{ny} \Mc^{jk}_{yx} \varphi^k_x\,.
\label{eq:symblock}
\end{eqnarray}
Of course we want to represent this transformation solely on the
lattice fields.  Indeed, the transformation can be lifted to $\phi$ as
\begin{eqnarray} 
 \phi\to\phi+\delta\phi\,,\qquad 
 (\de\p)_n^i=\epsilon \M_{nm}^{ij}\p^j_m\,,
 \label{eq:quantvar}
\end{eqnarray}
with a lattice transformation $\M$, if 
\begin{eqnarray}\label{eq:constraint0}
\M_{nm}^{ik} f_{mx}^{kj}= f_{ny}^{ik}\Mc_{yx}^{kj} \,. 
\end{eqnarray} 
holds.

This property can be viewed as a constraint ensuring the compatibility
of the lattice symmetry transformation with the blocking.  One might
be tempted to use it to define $\M$, but $f$ has no right-inverse,
since it maps onto fewer degrees of freedom.  This constraint has been
mentioned without further investigation in \cite{So:1998ya}.  We will
analyse it in full detail in Section~\ref{sec:constraint} and argue
that it has severe consequences, if $\Mc$ contains a derivative.
Since the transformations defined by $M$ act on lattice fields they
can be regarded as a naive realisation of the symmetry transformations
on the lattice.

According to eq.~\eq{eq:quantvar} 
this naive transformation $\delta$ has the operator representation 
\begin{eqnarray}  
  \delta =\epsilon \M_{nm}^{ij}\phi_{m}^j \0{\delta}{\delta 
    \phi_n^i}  
\label{eq:derrep_eff} 
\end{eqnarray} 
on the space of fields $\phi$. In the fermionic sector the left
derivative is used, e.g.\ $\frac{\delta}{\delta\psi}
\bar{\psi}\psi=-\bar{\psi}$. 

Now we can transform the classical symmetry into a relation of the
effective theory on the lattice.  To that end we apply in equation
\eq{eq:SWilson} the naive symmetry transformation $\phi\to \phi+\delta
\phi$ defined above to the Wilsonian action as well as the classical
symmetry transformation $\varphi\to \varphi +\dec\varphi$ to the
integration variable using that the invariance of the classical
action, $\delta \Sc[\varphi]=0$.  To linear order in $\epsilon$ we
arrive at
\begin{eqnarray} \nonumber &&\hspace{-.2cm} \M_{nm}^{ij}\phi_m^j
  \0{\delta }{ \delta \phi_n^i} \Sw[\phi] =-\Str\, \Mc - e^{\Sw[\phi]}
  \int d\varphi\,
  e^{-\Sc[\varphi]}  \\
  &&\hspace{.5cm} \times \M_{nm}^{ij}(\phi-\phi_f)^j_m \0{\delta
  }{\delta \phi_n^i} e^{-\012(\phi-\phi_f) \alpha (\phi-\phi_f) }\,.
 \label{eq:deltaWilson}\end{eqnarray} 
The supertrace term $\Str\,\Mc$ on the rhs comes from the expansion
of the Jacobi determinant and comprises the possible anomaly of the
symmetry transformation $\dec$.

The difference $(\phi-\phi_f)$ in this equation can be expressed as a
$\phi$-derivative using
\begin{eqnarray}
  \nonumber &&\hspace{-.3cm} (\phi-\phi_f)_m^j \0{\delta
  }{\delta \phi_n^i} e^{-\012(\phi-\phi_f) \alpha (\phi-\phi_f) }= -
  \Bigl( (-1)^{|\phi^i|} \delta_{mn} \delta^{ij}\\ &&\hspace{.5cm}
  +{\alpha^{-1}}_{mr}^{jk} \0{ \delta}{\delta \phi^k_r} \0{\delta
  }{\delta \phi^i_n}\Bigr)e^{-\012(\phi-\phi_f) \alpha (\phi-\phi_f) }
  \,.
  \label{eq:deltaphi-der} 
\end{eqnarray}
When inserted into \eq{eq:deltaWilson}, the first term on the rhs of
this identity contracts to the supertrace $+\epsilon\,\Str\,\M$ of the
lattice symmetry $\M$, while the rest only contains derivatives
w.r.t.\ the blocked field $\phi$ and can be pulled outside the $\vp$
integral. Then \eq{eq:deltaWilson} turns into
\begin{eqnarray} \nonumber &&\hspace{-1.3cm} \M_{nm}^{ij}\phi_m^j
\0{\delta \Sw[\phi] }{\delta \phi_n^i } = \Str \M-\Str\Mc \\ & & +
e^{\Sw[\phi]} ( \M \alpha^{-1})^{ij}_{nm} \0{ \delta}{\delta
\phi^j_m}\0{\delta }{\delta \phi_n^i} e^{-\Sw[\phi]}\,.
  \label{eq:deltaWilson1}
\end{eqnarray} 

Finally, performing the derivatives leads to a nonlinear relation for
the blocked action $\Sw$ containing up to second order derivatives in
$\phi$,
\begin{widetext}
\begin{eqnarray}
\M_{nm}^{ij}\phi_m^j \0{\delta \Sw
}{\delta \phi_n^i} =
(\M \alpha^{-1})_{nm}^{ij} \left( \0{ \delta
\Sw}{\delta \phi_m^j}\0{\delta \Sw }{\delta \phi_n^i}- \0{ \delta^2
\Sw}{\delta \phi_m^j\delta \phi_n^i}\right) 
+(\Str
\M-\Str\Mc)\,.
\label{eq:SymWilson}
\end{eqnarray}
\end{widetext} 
This is the key relation for the Wilsonian or lattice action
$\Sw[\phi]$, the naive symmetry transformation $\M$ and the blocking
kernel $\alpha$.  While the lhs.\ of this relation is just the naive
symmetry variation of the action $\Sw$, the rhs constitutes some
nontrivial modification of it, that has been derived in the blocking
procedure.  The behaviour of this term with respect to the continuum
limit will be investigated in section~\ref{sec:contlim}.  Note
furthermore that eq.~\eq{eq:SymWilson} represents the lattice version
of the (modified) quantum master equation, see e.g.
\cite{Pawlowski:2005xe,Litim:1998nf,Bonini:1993kt,Igarashi:2001mf,%
Barnich:2000zw}.

A few comments are in order here.  The supertraces of $\Mc$ and
$\M$-terms in this relation carry the non-invariance of the measures
$d\varphi$ and $d\phi$ respectively, and hence comprise possible
(integrated) anomalies of the theory.  More precisely, $\Str\Mc$
carries the full anomaly related to the measure $d\varphi$.  The
blocking removes a part of the integrations from the path integral
leaving only the field $\phi$ to be integrated.  The related part of
the anomaly leads to $\Str\Mc-\Str\M$.

The above derivation also works for a finite blocking step where
$\varphi$ is already a blocked field and $x,y$ are lattice
coordinates.  Then, however, the starting point must be regarded as a
Wilsonian action for a fine lattice ($\Sc[\vp]=\Sw[\vp]$) that already
satisfies the relation \eq{eq:SymWilson}, which e.g. brings in the
continuum anomaly.  This again leads to \eq{eq:SymWilson} for the
blocked effective action on the coarser lattice.

If the right hand side of \eq{eq:SymWilson} vanishes, we are left with
the invariance of the action under the naive symmetry transformations,
\begin{equation} 
  \M_{nm}^{ij}\phi_m^j \0{\delta \Sw}{\delta
    \phi_n^i} = 0\,.
  \label{eq:Sympre}
\end{equation}
This happens for symmetric blocking matrices $\alpha_S$
fulfilling
\begin{equation}
\label{eq:vanish} 
\M\alpha^{-1}_S\pm (\M\alpha^{-1}_S)^T=0 \,,
\end{equation} 
since only the (anti)symmetric part of $\M\alm$ enters the rhs of
the relation.
The minus sign appears only if the matrix $M$ connects fermions with fermions,
i.e.\  if fermionic fields are transformed into themselves by the symmetry.
The above condition just means that the blocking kernel is
invariant under the naive symmetry variation.  More generally,
$\alm+\alm_S$ leads to the same symmetry relation \eq{eq:SymWilson}
for all $\alm_S$. This defines a family of equivalent blocking kernels
$\alm(\alm_S)$.

For the chiral case $\al=\al_S$ is excluded by the vector symmetry
as we will elaborate in Section~\ref{sec:chiral}.  For the case of
supersymmetry, there is in general not such an argument and, indeed, 
such a matrix has been used e.g.\ in \cite{So:1998ya}.  However, we
shall show below, that the naive symmetry $M$ in the systematic
blocking approach to SUSY is inherently non-local and hence excluded.
 Instead, a non-symmetric blocking kernel
$\alpha$ must be used. 

Then the relevant symmetry can be written in terms of a modified
field-dependent symmetry operator, $\M_{\rm def}(\phi)$, defined as
\begin{equation} \label{eq:ModSym}
(\M_{{\rm def}})_{nm}^{ij}\phi_m^j :=
\M_{nm}^{ij}\left(\phi_m^j- (\alpha^{-1})^{jk}_{mr}\0{ \delta
\Sw}{\delta \phi_r^k} \right)\,. 
\end{equation} 
Inserting this definition into the symmetry relation
\eq{eq:SymWilson}, we are led to the relation
\begin{align} 
  (\M_{{\rm def}})_{nm}^{ij}\phi_m^j \0{\delta \Sw }{\delta \phi_n^i}&
  =(-1)^{|\phi^i||\phi^j|} \0{ \delta}{ \delta \phi_n^i}\! \left[
    (\M_{\rm def})_{nm}^{ij}\phi_m^j \right]\nonumber\\&\qquad\quad
  \qquad-\Str\Mc\,,
\label{eq:MdefSymWilson}
\end{align}
the right hand side being related to a total field-derivative. The
above derivation of $M_{\rm def}$ follows closely the analogous
continuum arguments as used in \cite{Pawlowski:2005xe,Litim:1998nf,%
  Bonini:1993kt,Igarashi:2001mf}. A discussion of various
representations of \eq{eq:MdefSymWilson} and their use can be found
in \cite{Pawlowski:2005xe}.

\subsection{Local lattice symmetries}
\label{sec:localityofMdef}

It is important to emphasise that \eq{eq:MdefSymWilson} in general
does not comprise a symmetry, as the above construction applies to any
blocking kernel $\alpha$.  Thus, in general \eq{eq:MdefSymWilson} only
disguises an explicit symmetry breaking induced by the blocking.
We shall exemplify this statement in section~\ref{sec:chiral} at the
standard Wilson Dirac operator that explicitly breaks chiral symmetry,
but still satisfies \eq{eq:MdefSymWilson}.

The question arises what are the additional conditions on $M_{\rm
  def}$ that make it to a deformed symmetry. For local continuum
symmetries it is important that the corresponding lattice version of
the symmetry carries this locality. More generally, for a given
symmetry the blocking should only induce a local symmetry breaking or
deformation generated by $f$ and $\alpha$. Consequently we are led to
two conditions:

\begin{itemize} 
\item[(1)] a mandatory condition for a deformed lattice symmetry is
  the locality of $M_{\rm def}$. This guarantees a well-defined
  continuum limit in which the lattice artefacts related to the
  deformation tend to zero in a controlled way as they are local.
  Hence, in order to have a deformed symmetry the family of blockings
  $\alpha^{-1}(\alm_S)$ must contain at least one blocking
  $\alpha^{-1}_{\rm local}$ that leads to a local symmetry operator
  $M_{\rm def}$. We emphasise that this does not necessarily imply
  that $\alpha^{-1}_{\rm local}$ is local. Locality of $M_{\rm def}$
  reads
\begin{equation}\label{eq:locality}
\lim_{|x-y|\to \infty}|M_{\rm def}(x,y) | < e^{- c |x-y|} 
\end{equation} 
for some $c>0$. In the present investigation we shall relax
\eq{eq:locality}, and demand
\begin{equation}\label{eq:genlocality}
  |x^r M_{\rm def}(x,y)|<\infty\quad \forall\, r \in 
  \mathbbm{N},\, x,y\in a\mathbbm{N}\, , 
\end{equation} 
for explanations see Appendix~\ref{app:local}. Clearly operators
$M_{\rm def}$ with \eq{eq:locality} satisfy \eq{eq:genlocality} but
\eq{eq:genlocality} also allows for softer decay, e.g. polynomial
times exponential decay. Moreover, for interacting theories the
locality conditions \eq{eq:locality},\eq{eq:genlocality} involve
field-dependent terms as $\M_{\rm def}(\phi)$ in \eq{eq:ModSym} is
field-dependent.

\item[(2)] $M_{\rm def}$ has to carry the original continuum symmetry
  related to the symmetry operator $\Mc$. This condition excludes
  e.g.\ the trivial solution $\M_{\rm def} \equiv 0$. For this
  solution it is clear that the symmetry pattern of the lattice action
  is not entailed in $\M_{\rm def}$, and $\M_{\rm def}$ does not tend
  towards the continuum symmetry $\Mc$ in the continuum limit. We can
  summarise this condition in the demand that $\M_{\rm def}$ is
  identical to the continuum symmetry operator $\M_{\rm cont}$ up to
  lattice artefacts at $p=0$, where the continuum limit is located.
  Hence, the condition that $\M_{\rm def}$ carries the continuum
  symmetry can be formulated as
 \begin{equation}\label{eq:mdef_contsym}
\lim_{p\rightarrow 0} M_{\rm def}=M_{\rm cont}(\id +O(ap))\,.
\end{equation}
Note that $M_{\rm cont}=\Mc$ only for $\alpha^{-1}=0$ in the
con\-tinuum, see the discussion in Section \ref{sec:contlimgen}.
\end{itemize}

The above conditions \eq{eq:ModSym}-\eq{eq:mdef_contsym} should be
seen as a definition of a deformed symmetry, and put constraints on
the blocking kernel $\alpha$.  In order to formally obtain a symmetric
continuum limit, one might use actions without such a symmetry, but
the above considerations guarantee the existence of a local lattice
symmetry for every finite lattice spacing that converges locally
towards the continuum symmetry.  The latter property is very important
for a successful numerical implementation.

In the case of lattice supersymmetry the question is whether 
such a deformed symmetry operator $M_{\rm def}$ 
according to this definition can be constructed.

\subsection{Quadratic action}
\label{sec:quadAction}

The general results above simplify greatly for quadratic actions
\begin{equation}\label{eq:quadratic} \Sw=\half\phi^i_n
K^{ij}_{nm}\phi^j_m\,,
\end{equation} with the kernel $K$ comprising kinetic and mass terms.
At first sight this case seems trivial as it describes a free field
theory.  Nonetheless, it already includes the non-trivial case of
Ginsparg-Wilson fermions \cite{Ginsparg:1981bj} with background gauge 
fields, see the next subsection. Moreover, locality of a symmetry operator 
$\M_{\rm def}$ of an interacting 
theory relates directly to the locality of its kinetic non-interacting part. 

With this action the general symmetry relation \eq{eq:SymWilson}
simplifies to
\begin{eqnarray}\nonumber && \hspace{-1.2cm} \phi \M^T K \phi = \phi\,
K^T (\M \alm)^T K\,\phi \\ && -\mathrm{tr}\, (\M \alm)\, K^T +(\Str
\M-\Str\Mc)\,.
\label{eq:quadGW}
\end{eqnarray} In many cases the second line vanishes. In case of
field independent transformation matrices $\M$ and kinetic operators
$K$ it anyway is just an irrelevant constant.  However, in the case of
anomalous symmetries it contributes to the anomaly. 
If one considers non-quadratic actions 
the corresponding term in general becomes
$\phi$-dependent.

The first line in \eq{eq:quadGW} has to be valid for general fields
$\phi$ and hence we conclude that
\begin{eqnarray}
\label{eq:matrix2}
 &&\M^T K \pm (\M^T K)^T \nonumber\\
&& = K^T (\M \alm)^T K\pm (K^T (\M \alm)^T K)^T\,.
\end{eqnarray}
Again the minus signs appear on the left and right hand side only if fermions are transformed into fermions by the naive symmetry $M$.

The interesting information in the symmetry relation is that of the
propagation of symmetry breaking on the lattice. 
This propagation can be seen from
\begin{equation}\label{eq:compensate}
 (K^{-1})^T M^T\pm MK^{-1}=(\alm)^T M^T\pm M\alm\, .
\end{equation}
This equation high-lightens how the breaking of the symmetry by
the blocking matrix $\alpha$ and the breaking by the kernel $K$ must
compensate each other. It also enables us to 
read-off the general solution $K$, 
\begin{equation}\label{eq:Ksol} 
K^{-1} =\alpha^{-1} -\alm_{S}\, .
\end{equation} 
Here, $\alm_{S}$ is a general symmetry-preserving term fulfilling
\eq{eq:vanish}.  We emphasise that \eq{eq:Ksol} can also be used for
determining a family $\alpha(K)$ for a given $K$. We conclude that
pairs $(K^{-1},\alm)$ are unique up to symmetry-preserving terms
$\alpha_S^{-1}$. The symmetry relation can be also rewritten by
introducing the deformed symmetry matrix $\M_{\rm def}$ as defined in
\eq{eq:ModSym}. Here we find a $\phi$-independent $M_{\rm def}$ with
\begin{eqnarray}
\label{eq:deformedsym} 
\M_{\rm def}:=\M\left(\id -
\alpha^{-1} K\right)=-M\alm_S K \,. 
\end{eqnarray} 
$M_{\rm def}$ may, however, now depend on background fields via $K$ and
$\alpha_S^{-1}$, e.g.\ link variables if $K$ is the Dirac operator.
Note that \eq{eq:deformedsym} defines a family of symmetry matrices,
as $\alm_S$ is a general symmetric matrix satisfying \eq{eq:vanish}.
For the modified symmetry the relation \eq{eq:matrix2} reads
\begin{eqnarray}
\label{eq:matrix2def} 
\M_{\rm def}^T K\pm (\M_{\rm
def}^T K)^T=0
\end{eqnarray} 
As already mentioned in the previous section, \eq{eq:matrix2def} in
general does not comprise a symmetry, as the above construction
applies to any kinetic operator. Thus, in general \eq{eq:matrix2def}
only disguises an explicit symmetry breaking induces by the blocking
kernel. We also clearly see the necessity of the second condition
\eq{eq:mdef_contsym}: for $\alpha^{-1}=K^{-1}$ we have
$\alpha_S^{-1}=0$ and hence $\M_{\rm def} \equiv 0$. Then the modified
symmetry relation \eq{eq:matrix2def} carries no information about the
symmetry at hand.

In turn, only $\M_{\rm def}$'s in \eq{eq:deformedsym} with
\eq{eq:matrix2def} and the locality and continuum limit properties
\eq{eq:genlocality} and \eq{eq:mdef_contsym} respectively define
deformed lattice symmetries. 
\subsection{Chiral symmetry}
\label{sec:chiral}

We shall first discuss the above construction and conditions at the
example of the chiral symmetry. Consider an action out of the field
multiplet of two fermionic fields: $\phi=(\psi,\bpsi^T)$. The related
kinetic operator is given in terms of the Dirac operator
\begin{eqnarray}
\label{eq:dirac} 
\frac{K}{a^d}=\left(\begin{array}{cc}
0 & {-{\cal D}^T}\\ {\cal D} & 0\end{array}\right)\,,
\end{eqnarray} 
and the action \eq{eq:quadratic} reads $\Sw=
a^d\,\bar\psi \cal D\psi$.  As explained above, in our units the
quantities relevant for the continuum must get additional factors 
of $a^d$ to account for the integral.  The continuum
action is invariant under symmetry transformations generated by
\begin{equation}
\label{eq:chiralM} \Mc \varphi=\left(\begin{array}{cc}
\gamma_5 & 0\\ 0 & \gamma_5^T
\end{array} \right)\left(\begin{array}{c} \psi\\ \bpsi^T
\end{array} \right)\, ,
\end{equation} 
with $\gamma_5^\dag=\gamma_5\,$. Since the transformation acts only
algebraically on spinor indices it is easy to fulfil the constraint
\eq{eq:constraint0}.  The naive transformation is just the same as the
continuum transformation. A general blocking matrix $\al$ carries the
fermionic anti-symmetry and reads:
\begin{equation}\label{eq:chiral_alpha} \frac{\al}{a^d}=\left(
    \begin{array}{cc}0&-\al_1^{T}\\\al_1&0
   \end{array}\right)\, , 
\end{equation} 
with a general $\alpha_1$.
Note that in order to get a real action 
both ${\cal D}$ and $\al_1$ must be hermitian.
 Inserting the kinetic operator
\eq{eq:dirac}, the chiral transformation matrix \eq{eq:chiralM}, and
the general blocking \eq{eq:chiral_alpha} into \eq{eq:quadGW} we are
led to
\begin{equation}\label{eq:GWgen} \{{\cal D}\,,\,\gamma_5\}={\cal
D}\{\gamma_5\,,\,\al_1^{-1}\} {\cal D}\, , 
\end{equation} 
which comes from the field dependent part of \eq{eq:quadGW}.  It can
be rewritten in terms of a deformed symmetry,
cf.~\cite{Luscher:1998pqa}, which is according to the general
definition of $\M_{\rm def}$ in \eq{eq:deformedsym} given as
\begin{equation}
\label{eq:tildeg} 
\M_{\rm def}= \left(\begin{array}{cc}
\defgamma  & 0\\ 0 & (\bardefgamma)^T
\end{array} \right)
\end{equation}
 with
\begin{eqnarray}\label{eq:defgamma}
\defgamma= \gamma_5(1-\al_1^{-1}\Di)\,,\qquad 
\bardefgamma= (1-\Di\al_1^{-1})\gamma_5 .
\end{eqnarray} 
In terms of the deformed $\gamma_5$'s the symmetry relation reads
\begin{equation}
 \bardefgamma\Di+\Di\defgamma=0\, .
\end{equation}
For hermitian $\alpha_1^{-1}$ and $\Di$ we arrive at
$\bardefgamma=\defgamma^\dagger$.

In case of a theory with vector-symmetry the blocking should respect
it.  Hence the simplest $\alpha_1$ is a fermionic mass term with mass
$1/a$, $\alpha_1=1/a \id$, where $\id$ is diagonal with respect to the
lattice sites and the identity in Dirac space. The result,
\begin{equation}\label{eq:GW} \{{\cal D}\,,\,\gamma_5\}=2a {\cal
D}\gamma_5 {\cal D}\, ,
\end{equation} 
is the Ginsparg-Wilson relation \cite{Ginsparg:1981bj}.  

The deformed symmetry operator $\M_{\rm def}$ 
from \eq{eq:tildeg}, \eq{eq:defgamma}
is local due to the
ultra-locality of $\alpha^{-1}$ and the locality of ${\cal D}$.
It should, however, be noted that $\defgamma$ is not normalised, $\defgamma^2
\neq \id$, and even vanishes at the doublers. We conclude that
$\defgamma$ does not define a chiral projection. Indeed no such
normalised $\defgamma$ can be constructed for a single Weyl fermion,
see \cite{Jahn:2002kg}, as a consequence of the Nielsen-Ninomiya no-go
theorem. In the given example the normalisation of $\defgamma$ fails
at the doublers, it is neither smooth nor local.

The part of equation \eq{eq:quadGW}, that is independent of $\p$,
carries the integrated chiral anomaly,
\begin{equation} \label{eq:chiralan} \Tr \gamma_5 {\cal D}+ (\Tr_{\rm
lattice} \gamma_5 -\Tr_{\rm cont} \gamma_5)=0\,.
\end{equation} 
This constrains the continuum regularisation in terms of the lattice
blocking and vice versa. The blocking matrix, \eq{eq:chiral_alpha},
was chosen to have vector symmetry. It is thus preserving the vector
symmetry of the kinetic term \eq{eq:dirac}. It follows that $\Tr
\gamma_5 \alpha^{-1} {\cal D}=2 \Tr \gamma_5 {\cal D}=(n_+-n_-)_{\rm
  lattice}$, where ${n_+}_{\rm lattice},{n_-}_{\rm lattice}$ are the
numbers of fermionic zero modes with positive and negative chirality
respectively related to the (background) gauge field on the lattice.
If we choose a blocking compatible with axial symmetry, the related
term would vanish. Then, however, vector symmetry is broken, and we
would loose (background) gauge symmetry.  This analysis is reflected
in the well-known fact that $\Tr_{\rm cont} \gamma_5$ is
regularisation-dependent. Choosing a vector symmetric regularisation
of the trace, e.g.
\begin{eqnarray}
\label{eq:vector} 
\Tr_{\rm cont}
\gamma_5:=\lim_{\epsilon\to 0}\Tr_{\rm cont} \gamma_5 e^{\epsilon
{\cal D}_{\rm cont}^2}\,,
\end{eqnarray} 
we are led to $\Tr_{\rm cont} \gamma_5= n_+-n_-$, where $n_+,n_-$ are
the numbers of fermionic zero modes with positive and negative
chirality respectively related to the (background) gauge field in the
continuum. In turn, an axially symmetric regularisation leads to
$\Tr_{\rm cont} \gamma_5=0$. The corresponding trace on the lattice
vanishes, $\Tr_{\rm lattice} \gamma_5=0$. Note that the sum of zero
modes on the lattice here comes from the functional relation involving
$\Sw$. In summary we conclude from \eq{eq:chiralan} that full chiral
symmetry in the presence of a background gauge field is maintained iff
the lattice gauge field permits the same difference of positive and
negative chirality zero modes.

As an example for an explicit breaking of chiral symmetry we consider
Wilson fermions with Dirac operator ${\cal D}_W$, 
\begin{equation}\label{eq:DW}
  a {\cal D}_W= i \gamma_\mu \sin(ap_\mu) +  r \sum_\mu (
  1-\cos(ap_\mu)) \,.
\end{equation}
In this case chiral symmetry is explicitly broken due to the
momentum-dependent Wilson mass. We start with the relation
\eq{eq:Ksol} for general Dirac operators ${\cal D}$.  The
corresponding blocking kernel (cf.\ \eq{eq:chiral_alpha}) is given by
\begin{eqnarray}
\label{eq:Wilsonblock} 
\alm_1={\cal D}^{-1} +\alm_{1,S}\,, 
\end{eqnarray}
The singularity of ${\cal D}^{-1}$ at the centre of the Brillouin zone
has to be removed from $\alpha_{1,S}^{-1}$ in order to guarantee the
continuum limit of $M_{\rm def}\to M_{\rm cont}$,
\eq{eq:mdef_contsym}. This is achieved with
 \begin{eqnarray}
\label{eq:projectdirac}
 \alpha_{1,S}^{-1}= \gamma_\mu \,\0{1}{d}
\,\tr\, \gamma_\mu {\cal D}^{-1}+\Delta\alm_{1,S}\,, 
\end{eqnarray} 
with $\tr\gamma_\mu\gamma_\nu=-d \delta_{\mu\nu}$, and $d$ is the
space-time dimension.  We conclude that $\alm_1$ is
given by
 \begin{eqnarray}
\label{eq:almW}
 \alm_1= \id\, \0{1}{d}\, \tr\, {\cal D}^{-1}-\Delta\alm_{1,S}\,. 
\end{eqnarray} 
with a scalar first term proportional to $\id$, and a symmetric
contribution $\Delta\alm_{1,S}$ proportional to $\gamma_\mu f_\mu(p)$.
Note that the first term cannot be changed by $\Delta\alm_{1,S}$ and
hence carries the unique information about the symmetry-breaking part
of the kinetic operator $K$. This part is the same for all members of
the family $\alpha(K)$ of blockings corresponding to a given $K$.
Restrictions on this part will hence constrain the class of possible
lattice actions.

For Ginsparg-Wilson fermions we have $\alpha_1^{-1}=a \id$ and
$\Delta\alm_{1,S}=0$, that is $\alpha^{-1}$ has no symmetric part.
For Wilson fermions the choice $\Delta\alm_{1,S}=0$ leads to a
non-local $\alpha_1^{-1}$: some higher derivative of $\tr\, {\cal
  D}_W^{-1}(p)$ is not bounded at the origin and this contradicts
locality, see Appendix~\ref{app:local}.  Furthermore this non-locality
cannot be changed by the symmetric term $\Delta\alm_{1,S}$ except in
one dimension.  We conclude that the blocking $\alm$ related to the
Wilson-Dirac operator is inherently non-local.  Still a priori this
does not entail that the corresponding $\M_{{\rm def}}$ is non-local.
However, the product of this $\alm$ (including $\Delta\alm_{1,S}$)
with the Wilson-Dirac operator is also inherently non-local and enters
$\M_{{\rm def}}$.  Actually, the non-locality of $\M_{{\rm def}}$ is
most easily seen from the non-locality of the left hand side of
\eq{eq:compensate} in a Taylor expansion about $p=0$. We conclude that
there is no deformed chiral symmetry operator $\M_{\rm def}$ for
Wilson fermions. 

We summarise that the lattice blocking induces the continuum
regularisation.  In turn, if we have chosen a specific continuum
regularisation, this restricts the lattice blocking compatible in the
continuum limit. We conclude this analysis with the remark, that an
analysis of chiral transformations $\psi\to (1\pm \gamma_5)/2\, \psi$
completely fixes the relations, as the related integrated anomaly is
independent of the regularisation. This is at the heart of the lattice
observations made in \cite{Hasenfratz:2007dp,Gattringer:2008je}.

\subsection{Explicit solution for a quadratic action}

For a quadratic action it is also possible to solve \eq{eq:SWilson}
for the effective action $\Sw$ explicitly.  Assuming
$\Sc[\varphi]=\half \varphi_x^i \tilde{K}_{xy}^{ij}\varphi_y^j$, the
lattice action $\Sw[\phi]=\half\phi K\phi$ can be obtained via
performing the Gaussian integration.  It leads to
\begin{equation}
 \label{eq:fixop}
 K=\alpha-\alpha f(f^T\alpha f+\tilde{K})^{-1}f^T\alpha .
\end{equation}
After some manipulations that can be found in appendix~\ref{sec:sfaqa}
the resulting fixed point operator reads in momentum space  
\begin{equation}
\label{eq:BietenK}
K(p_k)=\left(\sum_{l\in\mathbbm{Z}}\frac{f^\ast(p_k+l \frac{2\pi}{a} ) 
    f(p_k+l\frac{2\pi}{a})}{\tilde{K}(p_k+l\frac{2\pi}{a})}
  +\alpha^{-1}(p_k)\right)^{-1} .
\end{equation}
Note that such a solution of the Ginsparg-Wilson relation was already
mentioned in \cite{Ginsparg:1981bj}.  It is often called perfect
lattice action. 

In most cases $f(x)$ is considered to be the
averaging over one lattice spacing, e.g. in one dimension
\begin{equation}
\label{eq:flataverage}
 f(x)=\left\{ \begin{array}{l l} 
1/a &\mathrm{if}\; |x|<a/2\\
0 &\mathrm{otherwise}
\end{array}\right.\, ,
\end{equation}
which means $f(p_k)=\frac{2}{La}\frac{\sin(p_k a/2)}{p_k}$.
Such an averaging was applied in \cite{Bietenholz} to construct 
a free supersymmetric (perfect) lattice theory. 
However, since the constraint \eq{eq:constraint0} was not
considered there the symmetry properties of the resulting effective
action can not be expressed in terms of a lattice symmetry involving
only lattice fields:  
equation \eq{eq:constraint0} demands for the derivative operator
appearing in the supersymmetry transformations
\begin{equation}
\label{eq:bietensym}
 \sum_m\n_{nm}\phi(am)=\frac{1}{a}(\vp(an+a/2)-\vp(an-a/2))\, 
\end{equation}
and this can not be fulfilled for any $\n_{nm}$ since the transformation
involves the continuum fields.

To interpret the rhs of equation \eq{eq:bietensym} a new field was
introduced in \cite{Bietenholz}, which is defined to be
$\frac{1}{a}\vp(an+a/2)$ at the lattice point $an$.  Then the lattice
fields are transformed into such fields under the supersymmetry
transformations.  They are, however, rather a continuum than a blocked
lattice quantity.  The correct SUSY continuum limit is therefore
ensured in this approach because the lattice action is a direct
solution of the blocking.  But this property can not be expressed in
terms of a lattice symmetry, that contains only lattice fields.  A
well defined lattice symmetry is, however, desirable as a guiding
principle for the construction of a more general lattice action.

\section{Additional constraint}
\label{sec:constraint}

\subsection{Discussion}

In the derivation of the relation of the effective action there has
emerged a novel constraint, eq.~\eq{eq:constraint0},
\begin{eqnarray}\label{eq:constraint0_recap}
\M f= f\Mc \,, 
\end{eqnarray}  on the symmetries
$\M,\,\Mc$ and the averaging function $f$. It is trivially fulfilled
if the symmetry transformation merely acts on the multiplet indices,
e.g.\ with $\gamma_5$ in the chiral case.

\begin{figure}[b] 
\includegraphics[width=0.7\linewidth]{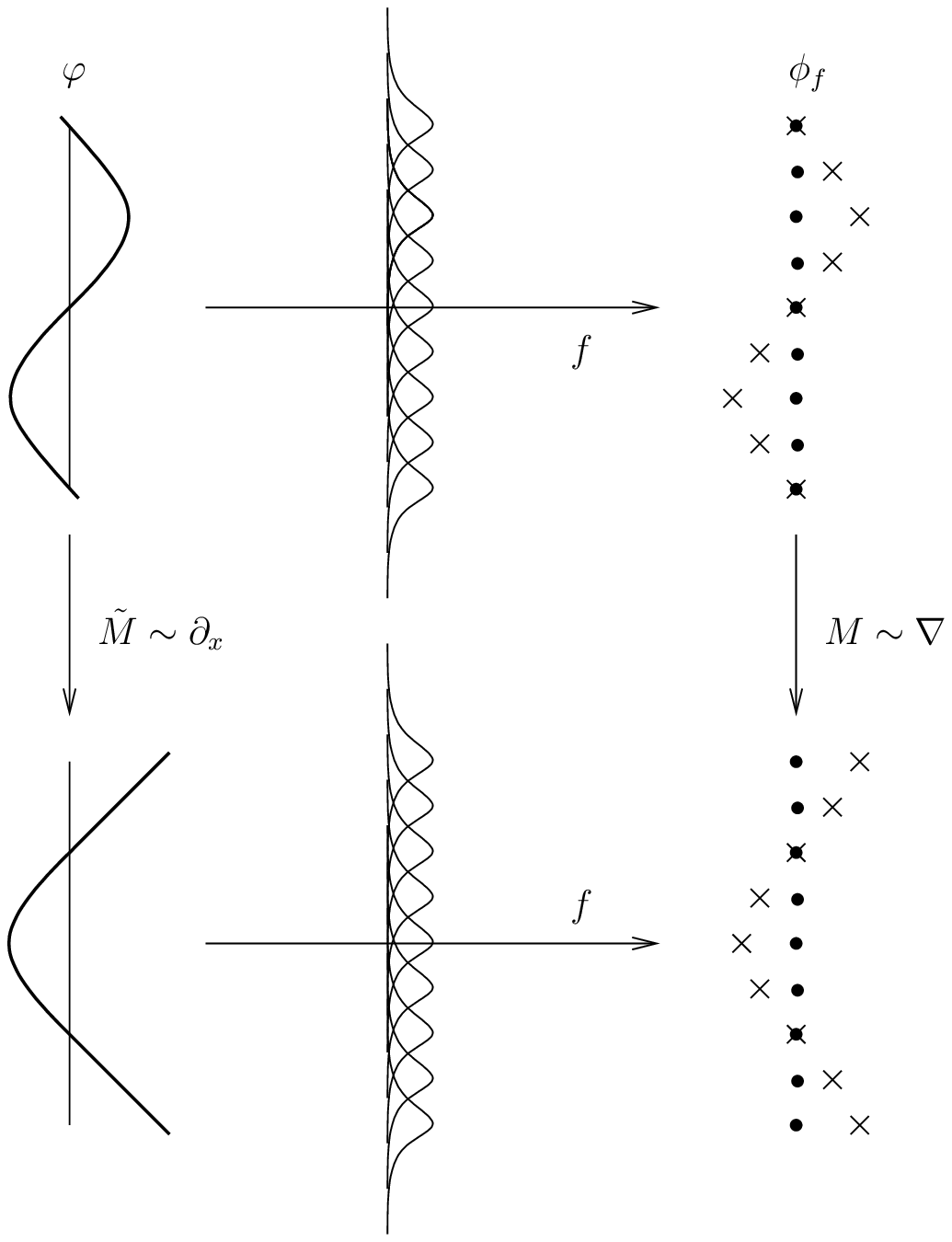}
\caption{A sketch of the blocking procedure: The averaging function
  $f$ maps from continuum fields $\varphi$ to averaged fields $\phi_f$
  (that are connected to the lattice fields $\phi$ via $\alpha$).  The
  additional constraint \protect\Eq{eq:constraint0} comes about
  because the diagram of $f$ with the continuum symmetry
  $\Mc\sim\partial_x$ and the lattice symmetry $\M\sim\n$ has to
  commute.}
\label{fig:add_constr}
\end{figure}

However, whenever the symmetry transformation $\Mc$ contains a
derivative -- as in the case of supersymmetry -- the constraint
becomes nontrivial. The problem can be considered in each space-time
direction separately.  It states, that the derivative $\partial$ (in
$\Mc$) is `pulled through' the averaging function $f$ to become a
lattice derivative operator $\n$ (in $\M$) that acts among the
averaged fields:
\begin{equation}
  \n_{nm} \int\! \! dx\;  f(am-x)\, \vp(x) = 
  \int\!\! dx\; f(an-x)\, \partial_x  \vp(x)
\label{eq:intconst}
\end{equation}
for all continuum fields $\varphi(x)$ (neglecting internal indices
$i,j$) and for all lattice points $n$.  This constraint will restrict
the possible lattice derivatives $\n$ to be used in the lattice
symmetry transformations $\M$ as we show now.

In order to satisfy hermiticity and translational invariance, $\n$
should be an antisymmetric circulant matrix
\begin{equation} 
\n_{nm}=\frac{1}{2a}\sum_{l=-(N-1)/2}^{(N-1)/2}
c_{l}\,\delta_{n-m ,-l}\, ,
 \label{eq:circulant}
\end{equation} 
with real coefficients $c_{l}$ fulfilling $c_{-l}=-c_l$.  For
simplicity we have specialised to an odd number $N$ of lattice points.
The Kronecker symbol $\delta$ on the rhs is periodic with periodicity
$N$.

We use a partial integration in \eq{eq:intconst} and a Fourier
transform with a discrete momentum $p_q=2\pi q/L$ with $q\in
\mathbb{Z}$ and a lattice volume $L=Na$ (for details see appendix
\ref{sec:FTconst}) to arrive at
\begin{equation}
  f(p_q)\big[\n(p_q)-i\, p_q\big]=0 \quad\forall \; q\in \mathbb{Z}\,.
 \label{eq:constraint_final}
\end{equation}
Hence, the constraint states that the averaging function can have
non-vanishing Fourier components $f(p_q)$ for each wave number $p_q$
for which the difference operator has the `ideal' continuum dispersion
relation $\n(p_q)=i p_q$. The latter condition means that the naive
translation
\begin{equation}
  \left.(\partial_x e^{i p_q x})\right|_{x=an}=
  \sum_m\n_{nm} e^{i p_q a m}
\end{equation}
holds for all lattice points $n$ for this wave number.

At this point let us stress that because of the periodicity
$\n(p_{q+N})=\n(p_q)$ the square bracket in eq.
\eq{eq:constraint_final} can vanish only once for every $q \mod N$,
for all other $q$ $f(p_q)$ has to vanish.  With regard to the correct
continuum limit of $\n$ we demand this to happen -- if at all -- in
the first Brillouin zone, thus
\begin{equation} f(p_q)=0 \quad \mbox{for } |p_q|>
  \frac{\pi}{a}(1-\frac{1}{N})\, .
 \label{eq:constraint_final_trivial}
\end{equation}
This limits the spatial resolution of $f$ to $|\Delta x|\sim a$ which,
however, is a natural scale in the blocking approach to a lattice (see
\eq{eq:flataverage} for comparison).

\subsection{Solutions}
\label{sec:solutions}

Inside the first Brillouin zone the constraint
\eq{eq:constraint_final} introduces a kind of uncertainty relation
between the averaging function $f$ and the lattice difference operator
$\n$. For instance, if one demands an ultra-local operator ranging
over only one neighbouring point, $\n$ will be proportional to the
symmetric difference,
\begin{equation} 
\n=c_1\,\n^{{\rm symm}},\qquad \n^{{\rm symm}}_{nm}=
\frac{1}{2a}(\delta_{n+1,m}-\delta_{n-1,m})\,.
\label{eq:n_symm}
\end{equation} 
The dispersion relation is in this case
\begin{equation} 
\n(p_q)=c_1\, \frac{i}{a}\sin\big(a p_q \big)\, ,
\end{equation} 
and the bracket in \eq{eq:constraint_final} vanishes for $p=1$ iff
\begin{equation} 
c_1=\frac{2\pi/N}{\sin(2\pi/N)}\,.
\label{eq:c_first}
\end{equation} The proportionality factor $c_1$ indeed approaches 1 in
the continuum limit and hence $\n$ approaches the continuum derivative
as $\n^{{\rm symm}}$ does. As a consequence, $f$ has only the lowest
(and zeroth) Fourier-components ($p_k=\{-2\pi/L,0,2\pi/L\}$):
\begin{equation} 
f(x)=f_0+f_1\cos(2\pi x/L)\,.
\label{eqn_f_symm}
\end{equation} 
Hence for this ultra-local difference operator the averaging function
$f$ is very broad as it probes the whole space $x\in[-L/2, L/2]$.

The solution allowing for the next ($p=2$) Fourier component in $f$
demands the difference operator to spread over at least nearest and
next-to-nearest neighbours with coefficients
\begin{equation}\label{eq:next_best1}
\n_{nm}=\frac{1}{2a}\Big(c_1(\delta_{n+1,m}-\delta_{n-1,m})+
c_2(\delta_{n+2,m}-\delta_{n-2,m})\Big)
\end{equation} with the solutions
\begin{eqnarray}\ c_1&=&\frac{2\pi}{N}\frac{2\sin(4
\pi/N)-\sin(8\pi/N)} {\sin^2(4\pi/N)-\sin(2 \pi/N)\sin(8
\pi/N)}\label{eq:next_best_c1}\\ c_2&=&-\frac{2\pi}{N}\frac{2\sin(2
\pi/N)-\sin(4\pi/N)} {\sin^2(4\pi/N)-\sin(2 \pi/N)\sin(8
\pi/N)}\label{eq:next_best_c2}
\end{eqnarray}
which correctly approach $4/3$ and $-1/6$ in the limit $N\to\infty$.  

One can proceed in this way.  The more Fourier coefficients are
included in the averaging function, the less localised gets the
derivative operator. In general, $\n$ needs to spread to the $n$th
neighbours to enable $n$ non-vanishing Fourier coefficients in
$f(p)\;$\footnote{In this way derivatives interpolating between the
  symmetric derivative and the SLAC derivative are constructed.}.

At the extreme, in order to make $f$ as narrow as it can get, all
Fourier components $f(p_q)$ (in the first Brillouin zone) are needed.
The constraint \eq{eq:constraint_final} then leads to the non-local
SLAC-operator by definition \cite{SLAC}.  The coefficients in this
case are
\begin{equation} c^{SLAC}_{l}=(-1)^{l}\frac{2\pi/N}{\sin(\pi l/N)}\,.
\end{equation}

All of these operators can in principle be used inside $\M$ to
generate the lattice supersymmetry transformations. But there are
additional restrictions on $f$ further reducing these possibilities.

\section{Continuum limit}
\label{sec:contlim}

The results of the last section~\ref{sec:constraint} necessitate a
careful investigation of the continuum limit. We have argued in
Section~\ref{sec:Wilson_blocking},\ that the averaging function $f$
needs to approach the delta-distribution in the continuum limit.
Hence, approaching this limit, more and more Fourier components
$f(p_q)$ are needed.  As a consequence of the additional constraint
\eq{eq:constraint_final}, the lattice derivative operator $\n$ agrees
in Fourier space with the SLAC derivative for the increasing number of
modes $p_q$ with non-vanishing $f(p_q)$. In other words, the difference
operator becomes more and more extended over neighbouring lattice
sites, while $f$ gets narrower. The more neighbours are included in
the lattice derivative $\n$, the more demanding numerical simulations
will become.  One is therefore tempted to use the most localised
solution from this constraint.

Moreover, if combined with an appropriate blocking kernel $\alpha$,
the numerical effort could be reduced further. Thus it is advantageous
to determine the general setting giving access to the full set of
allowed blockings $f$'s and, in particular, $\alpha$'s.

For that purpose we reconsider the full generating functional in the
presence of external sources. The physics of the blocked fields
$\phi_f$ with blocking $f$, \eq{eq:average}, is carried by general
correlation functions of this field, $\langle \phi_f(a n_1)\cdots
\phi_f(a n_r)\rangle$. These correlation functions are generated by
\begin{eqnarray}\label{eq:genfuncblock} Z_f[J]=\0{1}\Norm\int
d\varphi \, e^{-\Sc[\varphi]+ J \phi_f[\varphi]} \,. 
\end{eqnarray} 
The correlations functions of $\phi_f$ naturally live on a lattice
defined by $n\in N_1\times N_2\times\ldots\times N_d$ resulting from
the blocking $f$. 

\subsection{Continuum limit of the blocking $f$}
\label{sec:cont_f}

In the limit $\phi_f[\varphi]\to \varphi$ the correlation functions
$\langle \phi_f(a n_1)\cdots \phi_f(a n_r)\rangle$ tend towards the
continuum correlations functions.  Accordingly, the most restrictive
constraint coming from the comparison of the lattice observables with
their continuum counterparts is that $f$ must approach the delta
function in the continuum limit. Then the lattice theory resembles the
continuum up to minor modifications. As stated in the previous section
this means that equation \eq{eq:constraint_final} must be fulfilled
for an increasing number of modes.

Let us work on a lattice with $N$ points and let $\n$ be more
localised than the SLAC derivative, i.e.\ have an `ideal' dispersion
relation $\n(p_q)=i p_q$ up to some momentum $p_{\rm
  max}<\pi(1-1/N)/a$.  The corresponding $f$ has non-vanishing Fourier
components up to this momentum.  Doublers will appear in the spectrum
of such operators; they can be removed as shown in
App.~\ref{sec:app_doublers} within our solution for supersymmetric
quantum mechanics.  Analysing the consequences for $f$, however, we
will argue against these solutions in the following.

\begin{figure}[h] 
\includegraphics[width=0.7\linewidth]{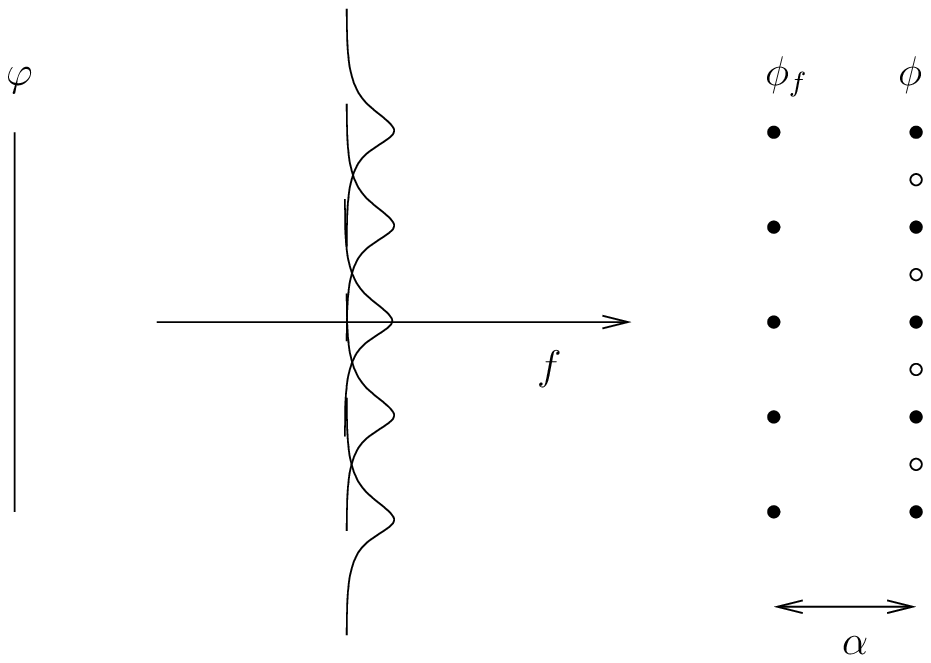}
\caption{A sketch of the mismatch in the blocking procedure, if $f$ --
  in order to generate a local lattice derivative in the constraint
  \eq{eq:constraint_final} -- has a limited number of Fourier
  components: the averaged fields $\phi_f$ have fewer degrees of
  freedom then the number of lattice points $N$ used for $\phi$. In
  other words, the $\phi_f$ transfer information from the continuum
  only to a coarser lattice (see text).}
\label{fig:mismatch}
\end{figure}

The momentum of the lattice fields $\phi(p_k)$ is restricted by this
momentum cutoff of $f$ (for an explicit formula see
\eq{eq:fourier_product}).  That means one introduces an additional
momentum cutoff smaller than the usual lattice cutoff.  In other words
the number of degrees of freedom (Fourier modes) induced from the
continuum via $f$ is smaller than the actual number $N$ of lattice
degrees of freedom.  This contradicts the blocking philosophy where
all the lattice degrees of freedom should come from blocked continuum
degrees of freedom.  To be more precise, in the defining equation
\eq{eq:SWilson}, the averaged fields $\phi_f$ span a vector space
smaller than the one of the lattice fields $\phi$.  Therefore some
lattice fields $\phi$ have no counterpart $\phi_f$.  Rather, their
contribution to the lattice action $\Sw[\phi]$ is a simple quadratic
one with kernel $\alpha$.  This mismatch is depicted in Fig.
\ref{fig:mismatch}.

Another way of stating the problem is that the blocking $f$ gives rise
to a resolution (an `effective lattice spacing') of $\pi(1-1/N)/
p_{\rm max}>a$.  On this coarser lattice the derivative is actually
again SLAC.  It is very unlikely that it yields any improvement to
work on the finer lattice with lattice spacing $a$, where not all of
the degrees of freedom are induced by a blocking from continuum
fields.

We conclude that the lattice derivative $\n$ entering the lattice
symmetry relation as $\M$ has to be the SLAC operator or some degrees
of freedom on the lattice will have no continuum counterparts.  In any
case relation \eq{eq:constraint_final} must hold for an increasing
number of lattice modes in the continuum limit if $f$ should approach
the delta distribution.

\subsection{Continuum limit of the generating functional}
\label{sec:contlimgen} 

It is left to discuss the consequences of a the general choice for
$\alpha$. This is best done in terms of the generating functional
$Z_f[J]$ defined in \eq{eq:genfuncblock}. 
The
path integral in \eq{eq:genfuncblock} can be conveniently rewritten in
terms of a path integral over lattice fields $\phi$. In
section~\ref{sec:Wilson_blocking} we have done this already for
vanishing external currents $J$, and a quadratic blocking kernel
$\012(\phi-\phi_f) \alpha (\phi-\phi_f)$, see \eq{eq:SWilson}, with
symmetry properties \eq{eq_alpha_symm} and \eq{fermbos}. This
procedure is readily extended to the general case with non-vanishing
currents by rewriting the source term $\exp \int J\phi_f$ via the
quadratic blocking kernel,
\begin{eqnarray}\label{eq:alpha} e^{J\phi_f}=\0{e^{-\s012
J\alpha^{-1} J}}{\int d\phi\, e^{-\012 \phi\alpha\phi}}\int d \phi\,
e^{-\012(\phi-\phi_f)\alpha(\phi-\phi_f)+J\phi}\,.
\end{eqnarray} 
Inserting \eq{eq:alpha} into \eq{eq:genfuncblock} the generating
functional $Z_f$ can be rewritten as a lattice generating functional
\begin{eqnarray}\label{eq:fullZ} Z_f[J] = \0{1}{\Norm[J]} \int
\d\phi\, e^{-\Sw[\phi]+ J\phi }\,,
\end{eqnarray} with Wilsonian action $\Sw$ as defined in
\eq{eq:SWilson}, 
\begin{eqnarray} \label{eq:SWilsonrecap} 
e^{-\Sw[\phi]}=\Sdet{}^{1/2}\alpha\, \int d\varphi\,
  e^{-\012(\phi-\phi_f) \alpha (\phi-\phi_f)-\Sc[\varphi]}\,. 
\end{eqnarray} 
The normalisation $\Norm[J]$ carries a trivial quadratic
dependence on the current $J$ and reads
\begin{eqnarray}\label{eq:norm} \Norm[J]=\Norm 
e^{\s012 J\alpha^{-1} J} \,. 
\end{eqnarray}
It reduces to $\Norm$ for vanishing current.  We emphasise again that
$Z_f[J]$ in \eq{eq:fullZ} has no dependence on $\alpha$, and reduces
to \eq{eq:Zphi} for vanishing current $J=0$. Note also that the
generating functional in \eq{eq:fullZ} with $\Norm[0]$ is the standard
lattice generating functional. As $\Norm[J]$ is a trivial Gau\ss ian,
lattice simulations for correlation functions straightforwardly
relate to those from \eq{eq:fullZ}.

The above construction allows us to evaluate general choices of
$\alpha$.  The blocking function $f$ will be discussed in the next
subsection, here we will assume $f$ to approach the
delta-distribution such that $\phi_f\to\varphi$. We have already
argued in section~\ref{sec:Wilson_blocking} that a diverging $\alpha$
ensures that the lattice field $\phi$ agrees with the blocked field
$\phi_f[\varphi]$. A simple $\alpha$ is a diagonal one,
$\alpha_{nm}^{ij}=c\delta^{ij}\delta_{mn}$ with $c\to \infty$, see
also section~\ref{sec:Wilson_blocking}.  Then we are lead to
\begin{eqnarray}\label{eq:deltaphirecap}
\Sdet{}^{1/2}\alpha\, e^{-\012(\phi-\phi_f)\alpha (\phi-\phi_f)}=
\delta(\phi-\phi_f)\,, 
\end{eqnarray}
see \eq{eq:deltaphi}, and the integral in \eq{eq:alpha} is trivially
done. Moreover, the normalisation looses its $J$-dependence,
$\Norm[J]\to \Norm$, as $\alpha^{-1}$ vanishes for $c\to \infty$.
Finally, the Wilsonian action is given by
\begin{eqnarray} \label{eq:SWilson0} 
e^{-\Sw[\phi]}=\int d\varphi\,\delta(\phi-\phi_f[\varphi])
  e^{-\Sc[\varphi]}\,. 
\end{eqnarray} 
This can be viewed as the canonical form of the Wilsonian action.
Note also that in this case the symmetry relation \eq{eq:SymWilson}
simplifies to the standard one, as the rhs vanishes with
$\alpha^{-1}=0$. 

In consequence the limit $\alpha^{-1} \to 0$ is the natural choice for
the continuum limit. In order to see how a general $\alpha$
scales with the lattice spacing $a$, we rewrite the blocking term as
\begin{eqnarray}
\label{eq:alphascalarrecap} 
&&\hspace{-1cm} (\phi-\phi_f)\alpha(\phi-\phi_f)\\ 
&&\hspace{.3cm}= a^d \sum_{i,\,n} a^d \sum_{j,\,m} 
(\phi-\phi_f)_n^i  \0{\alpha^{ij}_{nm}}{ a^{2d}} (\phi-\phi_f)^i_m \,.
\nonumber  
\end{eqnarray}
Firstly we remark that $a^d \sum_n \to \int d^d x$ in the continuum
limit $a\to 0$. Secondly, for symmetric smearings $\alpha$ we can 
always diagonalise the matrix $\alpha$. 
Thirdly, a factor $a^d$ appears because $\delta_{nm}/a^d\to\delta(x-y)$.
We conclude that all eigenvalues 
$\alpha_n$ have to satisfy 
\begin{eqnarray}\label{eq:alphacont} 
\alpha_n /a^d \to \infty\,, 
\end{eqnarray} 
for guaranteeing that the Wilsonian action tends towards the classical
action in the continuum limit. 

In other words, the inverse $\alm$ has to vanish for $\Sw\to \Sc$.
For practical purposes it might be advantageous to work with some
vanishing eigenvalues of $\alm$ already at finite lattice spacing.
Then the related eigenvalues $\alpha_n$ diverge already for finite
lattice spacing, and the right hand side of \eq{eq:deltaphirecap} will
be proportional to delta-functions for the related eigenfunctions
$\psi_n$, leading to $(\psi_n,\phi)=(\psi_n,\phi_f[\varphi])$.  We
also note that the symmetry relation \eq{eq:SymWilson} contains only
$\alm$, which shows no divergencies for $\alpha_n\to \infty$ but
zeros. Indeed we show in appendix \ref{sec:without_alpha}, that
\eq{eq:SymWilson} can be derived without using $\alpha$ explicitly,
and hence noninvertible $\alm$ are not problematic. On the contrary,
vanishing eigenvalues of $\alm$ mean that the factor $N[J]$ in
\eq{eq:norm} is actually a $J$-independent constant in that subspace
of fields.

Likewise, the eigenvalues of $\alm$ can have any sign
whereas \eq{eq:SWilsonrecap} would require positive eigenvalues of
$\alpha$.

What happens if some eigenvalues of $\alpha^{-1}$ do not vanish in the
continuum limit?  As the generating functionals $Z_f[J]$, and in
particular $Z[J]$, do not depend on $\alpha$, it is not mandatory that
the Wilsonian action $\Sw$ approaches the classical action $\Sc$.
Assuming $\phi_f[\varphi]\to\varphi$, a finite $\alpha$ in
\eq{eq:SWilsonrecap} amounts to equivalence classes of actions with
measures
\begin{eqnarray}
\label{eq:exponentials} 
d\phi\,\exp -\Sw[\phi]\,. 
\end{eqnarray}
related by Gaussian integrations in the continuum.  On the level of
classical actions this is nothing but the introduction of auxiliary
fields $\phi$ via the equations of motion.

The case of vanishing eigenvalues $\alpha_n$ (diverging $\alm$),
however, is different.  From the path integral in \eq{eq:SWilsonrecap}
one reads off that the corresponding subspace of fields $\varphi$ is
simply integrated out and the Wilsonian action $S$ does not depend on
the corresponding eigenfunctions $\psi_n$. The corresponding
singularities of $\alm$ in the normalisation \eq{eq:norm} can be
avoided by considering currents $J$ only in the orthogonal subspace,
that is $(\psi_n,J)=0$.  Then the above derivations are unaltered, but
with this procedure we have removed the $\varphi$-modes in the
singular subspace from our theory. A simple example for vanishing
$\alpha$ is given by the Wilson mass term in a fermionic theory.
Assume that we start with the naive lattice Dirac action with
doublers.  The blocking kernel $\alpha$ can be chosen such that it
vanishes at the doublers.  In turn $\alpha^{-1}$ provides a diverging
mass for the doublers.  More details on this and the general case with
vanishing $\alpha $ is deferred to Appendix~\ref{app:singalm}.

We summarise our findings as follows: in order to recover the original
action $\Sc=\Sw$ in the continuum limit $f\to \delta$, the blocking
kernel $\alpha$ has to lead to a delta-function in field space, see
\eq{eq:deltaphirecap}.  It might, however, be advantageous to rely on
a non-trivial classical Wilsonian action in the continuum limit,
generated by other choices of $\alpha$, in order to optimise the
locality and hence to minimise the numerical effort.

\section{Free supersymmetry on the lattice}
\label{sec:quadact}

We will now apply the blocking formalism to Supersymmetric quantum
mechanics (SUSYQM) which is a supersymmetric theory in one dimension,
i.e.\ all `fields' depend only on a time $x_1=t$.  It serves as a toy
model for supersymmetric theories.  After fixing the notation, only a
quadratic theory is considered in this section; interacting SUSY
theories will be considered in the next section.

\subsection{Brief review of SUSYQM in the continuum}

The field content of SUSYQM is the multiplet
\begin{equation}
 \varphi_x=\{\chi(t),F(t),\psi(t),\bpsi(t)\}
\end{equation}
where $\chi$ and $F$ are real bosons, $\psi$ is a complex fermion
(Grassmannian) and $\bpsi$ its complex conjugate. 
(The length dimensions of these fields are
$\sqrt{L}$, $1/\sqrt{L}$ and $L^0$, respectively.)

The Euclidean action in the continuum has the following form
\begin{equation}
  \Sc[\varphi]=\!\!\int\!\!  d t\Big[\half (\pa_t\chi)^2
  +\bpsi\partial_t\psi-\half F^2
  +\bpsi\, \frac{\partial W}{\partial \chi}\,\psi-FW(\chi)\Big]\,,
\label{eq:action_cont}
\end{equation}
where the first three terms are kinetic ones ($F$ is an auxiliary
nondynamical field) and the last two terms represent a potential term
for $\chi$ and Yukawa interactions.

This action is invariant under the supersymmetry transformations
\begin{equation}
\begin{array}{ccc}
  \delta\chi=-\bar{\ep}\psi+\ep\bpsi&\quad& \delta F=
  -\bar{\ep}\pa_t\psi-\ep\pa_t\bpsi\\
  \delta\psi=-\ep\pa_t\chi-\ep F&\quad&\delta\bpsi=
  \bar{\ep}\pa_t\chi-\bar{\ep} F .
\end{array}
\end{equation}
up to the following surface term
\begin{equation}
  \delta \Sc=\!\!\!\int\!\!dt\, \partial_t\!\left(\ep\bpsi(\pa_t\chi+F)
    +\ep \bpsi W(\chi)+\bar{\ep}\psi W(\chi)\right) .
\end{equation}

According to our general notation we write
\begin{equation}
  \delta \varphi = (\ep\Mc+\bar{\ep}\bar{\tilde{\mathcal{M}}}) \varphi 
\end{equation}
where
\begin{equation}
  \Mc
  =\left(\begin{array}{c c c c}
      0&0&0&1\\
      0&0&0&-\partial_t\\
      -\partial_t&-1&0&0\\
      0&0&0&0
\end{array}\right)
\quad
\bar{\tilde{\M}}
=\left(\begin{array}{c c c c}
    0&0&-1&0\\
    0&0&-\partial_t&0\\
    0&0&0&0\\
\partial_t&-1&0&0
\end{array}\right)\, .
\end{equation}

\subsection{Transformations on the lattice}
\label{sec:Mlat}

On the lattice it is very natural to take the same field multiplet,
now evaluated at discrete lattice points
\begin{equation}
  \phi_n=\{\chi_n,F_n,\psi_n,\bpsi_n\}
\end{equation}
In the corresponding lattice transformations (as defined by
\eq{eq:quantvar})
\begin{equation}
  \delta \phi_n^i = (\ep\M_{nm}^{ij}+\bar{\ep}
  \bar{\M}_{nm}^{ij}) \phi_m^j
\end{equation}
the matrices $\M$ and $\bar{\M}$ will be of the same form as in the
continuum,
\begin{eqnarray}
\label{eq:M}
\M_{nm}^{ij}
  &=&\left(\begin{array}{c c c c}
      0&0&0&1\\
      0&0&0&-\n\\
      -\n&-1&0&0\\
      0&0&0&0
  \end{array}\right)_{nm}\\
\bar{\M}_{nm}^{ij}
&=&\left(\begin{array}{c c c c}
    0&0&-1&0\\
    0&0&-\n&0\\
    0&0&0&0\\
    \n&-1&0&0
  \end{array}\right)_{nm} \,.
\label{eq:Mbar}
\end{eqnarray}
$\n$ is subject to the discussion in previous sections.  We will come
back to its locality properties in Section~\ref{sec:locality}.

\subsection{Ansatz for the quadratic lattice action}
\label{sec:quadr_susy}
For the rest of this section we restrict ourselves to a quadratic
theory.  Even in this simple case one obtains a nontrivial action with
nontrivial lattice derivative operators.  These operators solve the
relation \eq{eq:SymWilson} for SUSY just as the overlap operator
solves it for the chiral symmetry.

We choose the following matrix $K$ for the quadratic action (cf. sect.
\ref{sec:quadAction}):
\begin{equation}
\label{eq:K_SUSYQM}
\frac{K^{ij}_{mn}}{a}=
\left(
\begin{array}{c c c c}
  -\Box&-m_b&0&0\\
  -m_b&-\mathbbm{1}&0&0\\
  0&0&0&\tn-m_f\\
  0&0&\tn+m_f&0
\end{array}\right)_{mn},
\end{equation} 
which means
\begin{equation}
\label{eq:action_quadratic_SUSYQM}
\frac{\Sw[\phi]}{a}=-\half\chi\Box\chi+\bpsi(\tn+m_f)\psi-\half 
F\mathbbm{1} F
-Fm_b\chi\,.
\end{equation}
(again, a sum over the lattice indices is understood).  Though the
symbols of the matrices $K$ suggest them to be similar to the objects
in a quadratic continuum action with $W=m\chi$, they are so far
undetermined. (In particular the lattice masses and derivatives can be
different for fermions and bosons.)  $\Box$, $m_b$ and $m_f$ must be
symmetric and $\tn$ antisymmetric circulant matrices to guarantee
hermiticity and translational invariance.  In the continuum limit we
expect the behaviour $\Box\to\partial^2$, $\tn\to\partial$ and
$m_{b,f}\to m\mathbbm{1}$, while at finite lattice spacing these
matrices are chosen according to relation \eq{eq:SymWilson}. This also means that the matrix
$\tn$ can be different from the derivative operator $\n$ in the naive
generators $\M$ and $\bar{\M}$. Generalisations of the mentioned ansatz for the quadratic action are possible. One could, for example, introduce an additional
undetermined matrix in the $F^2$-term, but this is not necessary for
the solution of the relation in the next two sections. In the last section of this chapter we will consider the most general ansatz to derive  some statements for the general locality of the solutions.

\subsection{First solution for an ultralocal blocking}

To start with the simplest form of the blocking matrix $\alpha$ we
take its inverse to be
\begin{equation}
 \label{eq:redAlm}
 a(\alm)^{ij}_{mn}=
\left(
\begin{array}{cccc}
a_2 & 0 & 0 & 0 \\
0 & a_0 & 0 & 0 \\
 0 & 0 & 0 & a_1 \\
 0 & 0 & -a_1 & 0
\end{array}
\right)_{mn}\, ,
\end{equation}
where all $a_i$ are symmetric circulant matrices of length dimension
$i$ (and allowed to be zero, cf.\ Section~\ref{sec:contlimgen}).  
Note that these considerations also include a much more general ansatz. 
This happens because \eq{eq:redAlm} is the same as a general blocking ansatz
up to a symmetric part $\al_S$ as shown in appendix \ref{sec:GenAlpha}.

Circulant matrices commute. Using this property, a solution of the
symmetry relation \eq{eq:SymWilson} for the quadratic SUSY action \eq{eq:action_quadratic_SUSYQM} is
straightforward.  In Section~\ref{sec:quadAction} we have already
specified the symmetry relation for quadratic theories. From the first
line of \eq{eq:quadGW} one can read the following equations
\begin{eqnarray}
\label{eq:mrel1}
\!\!\!\!\! &&-\Box+\n\tn+\n (m_f-m_b)\nonumber\\
\!\!\!\!\! &&\;=-\Big[(a_2\n+a_1)\Box+(a_1\n+a_0)m_b\Big](\tn+m_f)\\
\label{eq:mrel2}
\!\!\!\!\! &&\tn-\n+m_f-m_b\nonumber\\
\!\!\!\!\! &&\;=-\Big[(a_2\n+a_1)m_b+(a_1\n+a_0)\Big](\tn+m_f)\, . 
\end{eqnarray}
Two additional equations can be identified with the transposed of
these if one reconsiders the symmetric or antisymmetric form of the
matrices.  The field-independent second line of \eq{eq:quadGW}
vanishes as $\M$ and hence $\M\alpha^{-1}K^T$ always connect bosons
with fermions.  The corresponding relation for the generator $\bar{\M}$
induces the same set of equations.

At first we proceed in the same manner as in the derivation of the Ginsparg-Wilson-relation: 
we use an ultralocal blocking with $a_i$'s diagonal in lattice sites 
and derive a solution for the lattice action in terms of these matrices. 
This solution corresponds to the overlap operator, 
that also is a function of the blocking (appearing in the chiral case).

  The second
equation \eq{eq:mrel2} can easily be solved for $\tn+m_f$ in terms of
$m_b$ and $\n$. One gets $\tn$ and $m_f$ as the antisymmetric resp.\
symmetric part of
\begin{equation}
 \frac{\Big[(1+a_1m_b+a_0)-(a_2m_b+a_1)\n\Big](\n+m_b)}
{X}\,,
\end{equation}
where
\begin{equation}
  X=(1+a_1m_b+a_0)^2-(a_2m_b+a_1)^2\n^2\,.
\end{equation}
The first equation \eq{eq:mrel1} then gives $\Box$ and the complete
solution reads
\begin{eqnarray}
  \!\!\!\!\tn&\!\!=\!\!&\frac{(1+a_0-a_2m_b^2)\n}{X}
  \label{eq:soln_tilde_nabla}\\
  \!\!\!\!m_f&\!\!=\!\!&\frac{(1+a_1m_b+a_0)m_b-(a_2m_b+a_1)
    \n^2}{X}\label{eq:soln_mf}\\
  \!\!\!\!-\Box+m_b^2&\!\!=\!\!&\frac{-\n^2+m_b^2}{1+a_0-a_2\n^2}
\label{eq:soln_boxmb}\,.
\end{eqnarray}
The last part is presented in terms of $-\Box+m_b^2$ because this operator appears in the bosonic sector after integrating out
$F$, i.~e.\ in the on shell action.

As expected, in the limit $a_i\to 0$ one has $\Box\to\n^2$, $
\tn\to\n$ and $m_f\to m_b$. At finite lattice spacing, however, these
operators are nontrivial because each of them contains both the
derivative operator $\n$ and the mass term $m_b$.  $\tn$ and $m_f$ are
nonsingular because $X$ is positive since $\n$ is antihermitian; $\Box$ is
nonsingular, if $a_0$ and $a_2$ are not largely negative and the
denominator in \eq{eq:soln_boxmb} vanishes. 

Now we have to check the locality of our resulting lattice action.
As explained in section \ref{sec:contlim}, the solution of the
additional constraint, \eq{eq:constraint0_recap}, leads to the
non-local SLAC derivative $\n(p)\sim p$.  
As the operators of the lattice
action \eq{eq:soln_tilde_nabla}-\eq{eq:soln_boxmb} are given in terms
of $\n$, one might expect that they inherit this locality problem.
Therefore, the locality properties of the lattice action need to be
examined carefully.

Since $\alpha$ is similar to a mass term (diagonal in lattice sites) the
form of the denominators in
\eq{eq:soln_tilde_nabla}-\eq{eq:soln_boxmb} renders the appearing operators
very similar to massive propagator.  In the continuum similar
expressions lead to an exponential decay for large distances.  On the
lattice, however, the corresponding behaviour is spoiled by terms
decaying only algebraically.  This is shown in appendix
\ref{app:propagator} using methods of complex analysis.

\subsection{Solution with a local action}

\label{sec:locality}
Since we insist on the locality of the lattice action,
more general blocking kernels $\alpha$ must be considered.
With these it is possible to enforce locality in the SUSY lattice action. 
In our point
of view this is a crucial feature of the modified symmetry including
the blocking kernel compared to the naive symmetry without it.

Allowing now for an arbitrary momentum dependence of $\alpha$ 
one can solve the equations
\eq{eq:mrel1} and \eq{eq:mrel2} for the circulant matrices $a_0$,
$a_1$ and $a_2$.  Consequently these matrices are dependent on $\tn$,
$\Box$, $m_f$ and $m_b$.  Given $\n(p_k)=ip_k$ the solution in terms
of the matrices of the lattice action is
\begin{eqnarray}
  a_0(p_k)&=& \frac{\Box}{m_b^2-\Box}-\frac{i \tn p_k}{m_f^2-\tn^2}
 \label{eq:a0}\\
  a_1(p_k)&=& \frac{-m_b}{m_b^2-\Box}
    +\frac{m_f}{m_f^2-\tn^2} \\
  a_2(p_k)&=& \frac{1}{m_b^2-\Box}+\frac{i \tn/p_k}{m_f^2-\tn^2}\, .
\end{eqnarray}
Now one could use the simplest ultra-local operators (without doublers)
in the action and read off the corresponding $\alm$ for a solution
of the relation. Additional restrictions for these operators
appear since singularities in the $a_i$ must be excluded.  A possible
singularity appears at $p_k=0$ if the mass is zero in the theory.
To avoid this singularity one can use $m_b=m_f$ and $\Box=\tn\tn$.
Note that in this way the fermionic and bosonic sector are treated in a similar manner. 
So possible doublers of the fermionic sector also appear in the bosonic one, 
but can be removed with the same mass term.
The result then simplifies to
\begin{eqnarray}
\label{eq:nonSingAlLocalK}
  a_0(p_k)=\frac{\tn(\tn-ip_k)}{m_f^2-\tn^2}\, , \quad 
  a_2(p_k)= \frac{1+i \tn/p_k}{m_f^2 -\tn^2}\, ,
\end{eqnarray}
and $a_1(p_k)=0$. Just as
expected all $a_i$ vanish if the derivative operator $\tn$ is the SLAC
derivative and the deformed symmetry is reduced to the naive one.

The simplest solution for the fermionic operators is the standard
Wilson fermion.  With the corresponding bosonic operators one finally arrives at
\begin{eqnarray}
 \label{eq:local_tilde_nabla}
&&   \tn(p_k)=\frac{i}{a}\sin(ap_k)\, ,\quad \Box=\tn\tn \\ 
 \label{eq:local_m}
&&   m_b(p_k)=m_f(p_k)=m+\frac{1}{a}\big(1-\cos(ap_k)\big)\, .
\end{eqnarray}
In this realisation all possible doublers are removed by the Wilson
mass terms in the bosonic and fermionic sector.

Such a form for the quadratic lattice action was chosen in
\cite{Catterall:2000rv} and other lattice simulations.  It has been shown
that the choice of the same mass and derivative operators in the
fermionic and bosonic sector leads to a major improvement with respect
to the lattice supersymmetry. 

Note, however, that a major requirement is still not considered. 
One should not only insinst on the locality of the action. 
To get a well defined representation of the symmetry on the lattice 
also the deformed symmetry transformations must be local.

\subsection{Local supersymmetry}
\label{sec:local_symmetry}

In order to get a well defined lattice supersymmetry, the deformed
symmetry operator $M_{\rm def}$ has to be local and approach the
continuum supersymmetry, as explained in Section \ref{sec:quadAction}.
The solutions we have obtained so far contain the SLAC operator,
either in the action, eqn.s
\eq{eq:soln_tilde_nabla}-\eq{eq:soln_boxmb}, or in the inverse
blocking kernel, eqn.s \eq{eq:a0}-\eq{eq:nonSingAlLocalK}.  This leads
to a non-local behaviour in $\alm K$, which in $M_{\rm def}=M-M\alm K$
can enhance or reduce the non-locality of $M$ (whereas in the chiral
case a non-local $\alm K$ immediately induces a non-local $M_{\rm
  def}$).  Here we investigate the conditions under which the locality
of the action and the deformed symmetry generator $M_{\rm def}$ can be
achieved.
We first derive a special solution which fulfils the locality condition \eq{eq:genlocality} for the action and $M_{\rm def}$. At the end of the section we will argue that this condition (and not \eq{eq:locality}) is the best one can achieve in the present setup.

For the special solution we consider now a slightly generalised form of the deformed supersymmetry and the
lattice action.  Since the corresponding inverse blocking kernel has
no direct physical implication it is adjusted accordingly.

In the ansatz for the quadratic lattice SUSY action, eq..\
\eq{eq:K_SUSYQM}, a general symmetric circulant
matrix $-I$ is used in the $F^2$-term instead of the diagonal
matrix $-\mathbbm{1}$. As a deformed supersymmetry generator we
take
\begin{equation}
 M_{\rm def}=
\left(\begin{array}{c c c c}
0&0&0&I_b \\
0&0&0&-\nabla_b\\
-\nabla_f&-I_f&0&0\\
0&0&0&0
\end{array}\right)\, . 
\end{equation}
If $\alm$ vanishes in the continuum limit, it follows that $M_{\rm
  def}\to \tilde{M}$, and hence $I_{b,f}\to 1$ and
$\nabla_{b,f}\to\partial_t$.

Demanding that $M_{\rm def}$ is a symmetry of the lattice action $K$,
eqn.~\eq{eq:matrix2def}, one arrives at the following sets of equations:
\begin{eqnarray}
 \frac{I_b}{I_f}&=&\frac{m_b}{m_f}=\frac{\nabla_b}{\nabla_f}\\
 \frac{\nabla_f}{I_b}&=&\frac{\Box}{\tn}\\
 \frac{\nabla_b}{I_b}&=&\frac{\tn}{I}\, .\label{eq:def_symm_3}
\end{eqnarray}
This can be solved easily, e.g.\ for the choice \eq{eq:local_tilde_nabla},
\eq{eq:local_m} of straightforward operators in $K$ one obtains $I=1$. 
In that case a natural solution is
\begin{eqnarray}
I_b=I_f=1,\quad \nabla_b=\nabla_f=\tn\, .
\end{eqnarray}
With $\tn$ as defined in \eq{eq:local_tilde_nabla}), for example, 
the deformed symmetry is ultra-local and obviously approaches the
continuum supersymmetry.  So it seems that there exists a local
deformed symmetry of the considered local lattice action.  The reason
for the absence of locality problems is that the SLAC operator from
the blocked lattice symmetry $M$ has not appeared yet.

The deformed symmetry must, however, not only be a local symmetry of
the lattice action. A further condition, eq.\ \eq{eq:deformedsym},
implies a relation between $M$ and $M_{\rm def}$. 
This additional condition is a direct consequence of the symmetry relation \eq{eq:SymWilson}.  Since $\alm$ is so
far undetermined it is apparently not difficult to satisfy this
condition.  But restrictions of the lattice action, due to e.g.\
hermiticity, impose further constraints on the blocking kernel.  In the case
of supersymmetry these constraint are of great importance since $\alm$
can only connect fermions with fermions whereas $M_{\rm def}$ connects
fermions and bosons with each other.  So one has to investigate
whether or not there exists an $\alm$ to ensure that this deformed
symmetry $M_{\rm def}$ is indeed the result of a blocking procedure
and hence fulfils $M_{\rm def}=M(\id - \alm K)$.   As $M$ is not invertible
we use a general ansatz for $\alm$, namely
\begin{equation}
 \label{eq:genAlm}
 a(\alm)^{ij}_{mn}=
\left(
\begin{array}{cccc}
  \elm_2 & \elm_1' & 0 & 0 \\
  \elm_1' & \elm_0 & 0 & 0 \\
  0 & 0 & 0 & \elm_1+\elm_1'' \\
  0 & 0 & -\elm_1+\elm_1'' & 0
\end{array}
\right)_{mn}\,,
\end{equation}
and compare the two sides of $M\alm=(M-M_{\rm def})K^{-1}$ yielding
\begin{eqnarray}
 &&\left(\begin{array}{c c c c}
0&0&(-1)(\elm_1-\elm_1'')&0 \\
0&0&\nabla_b(\elm_1-\elm_1'')&0\\
-\nabla \elm_2-\elm_1'& -\nabla \elm_1'-\elm_0&0&0\\
0&0&0&0
\end{array}\right)\nonumber\\
 &&=\left(\begin{array}{c c c c}
0&0& (1-I_b)/(\tn-m_f)&0 \\
0&0& (\nabla_b-\nabla)/(\tn-m_f)&0\\
\ldots&\ldots&0&0\\
0&0&0&0
\end{array}\right)\, .
\end{eqnarray}
The lower left block of this matrix equation can be satisfied 
by an appropriate choice of $\elm_0$, $\elm_1'$ and $\elm_2$.
The upper right block fixes $\elm_1-\elm_1''$ and in addition implies
\begin{equation}
 \frac{\nabla_b}{I_b}=\nabla\,.
\end{equation}
Together with \Eq{eq:def_symm_3} this yields
\begin{equation}
 I\,\nabla = \tn\, .
\end{equation}
We remind the reader that $\nabla$ is the SLAC derivative appearing in
the lattice symmetry $M$ due to the additional constraint and $I$ and
$\tn$ are the matrices for the $F^2$ and the kinetic fermion term in
the lattice action, which both should be local.

The SLAC operator is non-local because of its discontinuity at the
boundary of the Brillouin zone.  To ensure a local $\tn=I\,\nabla$,
i.e.\ to make it periodic in momentum space, $I$ and all its
derivatives must vanish at the boundaries of the Brillouin zone.  Then
the behaviour is not analytic in momentum space, but a stronger than
polynomial decay is guaranteed (see appendix \ref{app:local}).  
With such a result for the matrix $I$ all
symmetry conditions for the action can be fulfilled with
\begin{eqnarray}
 && m_f=m_b\,,\quad I_f=I_b=I\\[2ex]
 && \n_f=\n_b=\tn=I \n\,,\quad\Box=I \n^2\, .
\end{eqnarray}
All of these operators of the action satisfy locality stronger than polynomial.
Note that this solution amounts to $M_{\rm def}=I M$.

The only remaining problem arises since $I$ vanishes at the boundaries
of the Brillouin zone.  This means that the on-shell bosonic mass term
inherits a divergence $m_b(p)^2/2 I(p)$ at the edge of the Brillouin
zone.  Hence an on-shell problem is expected although the off-shell
action is local in the sense of condition \eq{eq:genlocality}.

It is instructive to recall that these results rely on the specific
structure of the transformations in the case of supersymmetry.  They
transform fermions into bosons and vice versa.  By contrast the
blocking matrix and the action relates fermions with fermions and
bosons with bosons.  In case of other symmetries the transformations,
$\alm$, and $K$ would have the same block diagonal structure, which
makes it easier to find a local solution for $K$ and $M_{\rm def}$.
 
One may still wonder whether the current approach can be generalised to get a result that satisfies the more severe condition \eq{eq:locality} for locality.
To investigate this problem we look at the relation $M_{\rm def}=-M\alm_S K$ 
(see \eq{eq:deformedsym}),
where only the symmetric part of $\alm_S$ of the blocking appears.
To get an exponential decay for both $M_{\rm def}$ and the action $K$,
also $M\alm_S$ has to fulfil this condition.
The SLAC-derivative
in the generator $M$, \eq{eq:M}, violates the locality of this matrix.
Taking the most general $\alm_S$, \eq{eq:MalmS}, the only entries of $M\alm_S$ are  $\sy{\elms}+\asy{\elms}$ ($\sy{\elms}-\asy{\elms}$ for the second supersymmetry, $\bar{M}\alm_S$) and the same matrix multiplied by the SLAC-derivative.
So both $\sy{\elms}+\asy{\elms}$ and $\nabla(\sy{\elms}+\asy{\elms})$ must be local. 
In analogy to $I$ and $I\,\nabla$ we conclude that it is impossible to get an exponential decay on the lattice for both matrices. So the condition \eq{eq:genlocality} and not \eq{eq:locality} can be fulfilled for both $M_{\rm def}$ and $K$. 

\section{Towards interacting supersymmetry on the lattice}
\label{sec:interacting}

The final goal of our investigations of lattice SUSY is to write down
a lattice action for an interacting theory.  According to the usual
argument, that is not blocking inspired, a supersymmetric lattice
action for the quadratic case can be found.  The lattice supersymmetry
transformations are in that case defined as the continuum
transformations with the derivative operator replaced by a local
lattice derivative.

For the case of a quadratic action we have given a solution above,
that preserves the modified symmetry and is hence guided by the
blocking of the symmetry.  In this case it is also possible to derive
a direct solution of the blocking transformations, cf.
eq.~\eq{eq:fixop}.  Therefore it is desirable to extend the above
results to interacting theories.

For actions beyond second order we go back to the original equation
\eq{eq:SymWilson} which provides a systematic relation to be fulfilled
in order to keep the considered symmetry.  The solutions are, however,
much more complicated than in the quadratic case.  This can already be
observed by an analysis of a specific solvable example, namely
constant fields in SUSYQM, which gives nontrivial results and displays
a new issue: the polynomial nature of the action.

\subsection{Solution for constant fields in SUSYQM}
\label{sec:constantfields}

In this subsection we work with the same parametrisations for $\M$,
$\bar{\M}$ and $\alm$, see \eq{eq:M} and \eq{eq:redAlm}, but study
only constant fields $\chi_n=\chi$ and so on.  This amounts to
the zero mode sector of the lattice derivative $\n$, i.e.\ we can
replace $\n\to0$.  Note that in this approximation $F$ is invariant
under the naive lattice transformations.  Of course, locality will not
be an issue for this toy model.

For the action we use the following ansatz:
 \begin{equation}
   \frac{S}{a} = N[\bp\psi \,g(\chi)-h(\chi,F)]
\end{equation}
where $N$, the number of lattice sites, is a remnant of the summation.
In view of the continuum limit we have restricted ourselves to an
$F$-independent function $g$ coupling the boson and the fermion, like
$\partial W/\partial\chi$ in the continuum action \eq{eq:action_cont}.
Likewise, we expect the $Fg=\partial h/\partial\chi$ to hold in the
continuum limit.

For such an action the relation \eq{eq:SymWilson} becomes 
a partial differential equation in $g$ and $h$:
\begin{equation}
  Fg-\frac{\partial h}{\partial\chi}=-\Nt a_1 g 
  \frac{\partial h}{\partial\chi}-\Nt a_0g\frac{\partial h}{\partial F}
  -\frac{a_1}{a}\frac{\partial g}{\partial\chi}\, .
\label{eq:eqn_const_fields}
\end{equation}
This indeed approaches $Fg=\partial h/\partial\chi$ in the 
limit where the $a_i$ vanish. 

For finite $a_0$ and $a_1$ this equation can be solved for $h$ for
different choices of $g$.  Among these solutions we consider those,
that consist of a term $F^2/2$ plus terms linear in $F$ with an
arbitrary $\chi$-dependence, such that the auxiliary field $F$ can be
integrated out easily.

The simplest case is $g(\chi)=0$, which should include the `kinetic
term', that in the zero mode sector degenerates to $-F^2/2$. Indeed,
any function $h(F)$ is a solution of eq.~\eq{eq:eqn_const_fields} in
this case.

The case $g(\chi)=m_f$ resembles an additional mass term. 
The corresponding solution
\begin{multline}
h(\chi,F)=\half F^2+\frac{1+a_0 \Nt}{1-a_1 \Nt m_f}m_fF\chi\\
+\frac{a_0}{2}\frac{ (1+a_0 \Nt)\Nt m_f^2}{(1-a_1 \Nt m_f)^2}\chi^2\, .
\end{multline}
becomes $F^2/2+m_fF\chi$ in the limit $a_i\to 0$ as required.  For
finite $a_i$ we obtain a bosonic mass different from $m_f$ and an
additional term proportional to $\chi^2$.  This action could also be
gotten from Section~\ref{sec:quadr_susy}, eqn.s
\eq{eq:action_quadratic_SUSYQM} and
\eq{eq:soln_tilde_nabla}-\eq{eq:soln_boxmb}, since putting $\n\to 0$
there also yields different bosonic and fermionic mass and a
$\chi^2$-term surviving from $\Box$.

The most interesting case is an interacting theory with a truly
$\chi$-dependent term $g(\chi)$. For the lowest possible power
\begin{equation}
 g(\chi)=\lambda\chi\,,
\end{equation}
one obtains the common Yukawa interaction.  The general solution of
\eq{eq:eqn_const_fields} is again restricted by the requirement that
for vanishing constants $a_i$ the term $h$ should resemble the
continuum result $F^2/2+\lambda F \chi ^2/2$. 
One obtains the non-polynomial solution
\begin{multline}
  h(\chi,F)=\frac{1}{2}{F^2}-\frac{1+a_0 \Nt }{a_1 \Nt}\chi F+
  \frac{a_0 (1+a_0 \Nt) }{2 a_1^2 \Nt}\chi^2\\
  - \left(\frac{1}{a\Nt}+\frac{1+a_0 \Nt}{a_1^2\lambda \Nt^2}F-
    \frac{a_0 (1+a_0 \Nt)}{a_1^3\lambda \Nt^2}\chi\right)
  \log (1-a_1\lambda \Nt \chi) \\
  +\frac{a_0 (1+a_0 \Nt)}{2 a_1^4\lambda^2 \Nt^3}(\log (1-a_1\lambda
  \Nt \chi))^2\, ,
\label{eq:h_full}
\end{multline}
which upon expanding in $a_1$ becomes polynomial
\begin{multline}
  h(\chi,F)=\frac{1}{2}F^2+\frac{\lambda}{2} (1+a_0 N) F \chi^2\\
  +\frac{a_0}{8}\lambda^2N(1+a_0 N)\chi^4 +O(a_1)\, .
\label{eq:h_series}
\end{multline}
Now it is obvious that the correct continuum limit is approached and
that this interaction term does not diverge in the limit $a_1\to 0$ as
one might have expected from \eq{eq:h_full}.

Note that in the zero mode sector there is no failure of the Leibniz
rule.  Hence the direct translations of the continuum action are
naively supersymmetric fulfilling \eq{eq:Sympre} which amounts to
setting $a_{1,2}=0$ (in other words using a SUSY preserving blocking
$\alpha_S$).  The presented solutions are deformations thereof.
Already in the simple case of constant fields these deformations lead
to a non-polynomial solutions if an interacting theory is considered.

One might expect that the non-polynomial form of the action is a
generic consequence of the symmetry relation \eq{eq:SymWilson}, not
only for supersymmetry. Indeed, similar results have been obtained in
the context of chiral symmetry \cite{Igarashi:2002ba}. We will discuss
this in the next sections as well as the circumstances under which a
polynomial action like \eq{eq:h_series} is possible.

\subsection{Polynomial solutions of the 
  symmetry relation}

As shown in the previous subsection it seems difficult to find a
polynomial solution of the lattice symmetry relation \eq{eq:SymWilson}
beyond second order.  In the following we argue that a non-polynomial
action is indeed the generic solution of this relation for an
arbitrary linear symmetry.  Only if special conditions are fulfilled,
the series in the fields can be truncated.  To show this general
behaviour we generalise the considerations of section
\ref{sec:quadAction} to interacting systems.

We consider a lattice action consisting of polynomials up to order $R$
in the fields represented as
\begin{multline}
\label{eq:actionseries}
\Sw[\phi]=\sum_{r=1}^R s^{(r)}[\phi] \,,
\;\; s^{(r)}[\phi]=K^{i_1\ldots i_r}_{n_1\ldots n_r}
\phi_{n_1}^{i_1}\ldots\phi_{n_r}^{i_r}\, ,
\end{multline}
where $s^{(r)}$ contains the $r$th order in the fields.  The purely
quadratic case $R=2$ (and $s^{(1)}=0$) has been discussed in section
\ref{sec:quadact}.  The coefficients $K$ are so far not further
specified, they can imply a simple multiplication of fields at the
same lattice point, but are also allowed to contain lattice
derivatives or to smear the powers of the fields over several lattice
sites, as long as they obey the correct continuum limit.

The relation \eq{eq:SymWilson} is in general a complicated nonlinear
differential equation coupling derivatives with respect to the fields
at different lattice points.  An expansion in the order of the fields
using the ansatz \eq{eq:actionseries} yields
\begin{widetext}
\begin{eqnarray}
  O(\phi^0):\qquad\qquad\,0&=&\M\alm\,\Big(\frac{
    \delta s^{(1)}}{\delta \phi}\frac{\delta s^{(1)}}{\delta \phi}-
  \frac{\delta^2 s^{(2)}}{
    \delta \phi\delta\phi}\Big)+(\Str \M-\Str\Mc)
\label{eq:General0}\\
  O(\phi^{r=1\ldots R-2}):\,\: \M\phi\,
  \frac{\delta s^{(r)}}{\delta \phi}&=&
  \M\alm\sum_{s+t=r+2}\Big(\frac{\delta s^{(s)}}{
    \delta \phi}\frac{
    \delta s^{(t)}}{\delta \phi}
  -\frac{\delta^2 s^{(r+2)}}{\delta \phi\delta\phi}\Big)
  \label{eq:General1}\\
  O(\phi^{r=R-1, R}):\,\:\M\phi\,\frac{\delta s^{(r)}}{\delta \phi}&=&
  \M\alm\sum_{s+t=r+2}\frac{\delta s^{(s)}}{\delta 
    \phi}\frac{
    \delta s^{(t)}}{\delta \phi}\label{eq:General2}\\
  O(\phi^{r=R+1\ldots 2R-2}):\qquad\qquad\,0&=&\M\alm
  \sum_{s+t=r+2}\frac{\delta s^{(s)}}{
    \delta \phi}\frac{\delta s^{(t)}}{\delta \phi}
  \,,\label{eq:General3}
\end{eqnarray}
\end{widetext}
where we used a short-hand notation without indices.
These coupled equations can be read as restrictions 
for the $K^{i_1\ldots i_r}_{n_1\ldots n_r}$ 
parametrising $s^{(r)}$ imposed by the symmetry. 
In the case of $R$\nolinebreak$=$\nolinebreak$2$ 
only the conditions \eq{eq:General0},\eq{eq:General1} are relevant
giving \eq{eq:quadGW}.

For interacting theories, $R>2$, a set of equations with vanishing
left hand sides, eqn.s \eq{eq:General3}, appears.  The very difficulty
to obtain a polynomial solution is to fulfil these equations with a
finite number of interacting terms $s^{(r)}[\phi]$ (which in addition
shall give the desired theories in the continuum).  If this turns out
to be impossible for some order $R$, this order has to be increased
and finally one might be forced to non-polynomial interactions.

As an example we consider the relation of the highest order
$O(\phi^{2R-2})$, which reads
\begin{eqnarray}
  0=(\M\alm)^{ij}_{nm}\Big(\frac{\delta s^{(R)}}{
    \delta \phi^j_m}\frac{\delta s^{(R)}}{\delta \phi^i_n}\Big)\,,
\end{eqnarray}
and can be rewritten in matrix-vector notation as
\begin{equation}
  0=v^T(\M\alm)\,v  \quad \mathrm{with}\quad v^i_n=\frac{\partial 
   s^{(R)}}{\partial \phi^i_n}\,.
\label{eq:vMv}
\end{equation}
This relation is a severe constraint, because it implies that
\begin{equation} 
\M\alm+(\M\alm)^T =0
\label{eq:truncationconst}
\end{equation}
within the subspace of lattice fields spanned by the $v^i_n$.
If the $v^i_n$ span the whole space of $\phi^i_n$,
the relation is immediately reduced to the naive symmetry.

If the $v^i_n$ do not span the whole space of the fields 
they must be linearly dependent. 
Then some linear combinations of the $v_n^i$ vanish and from 
the definition \eq{eq:vMv} it is clear that the highest part of the action 
$s^{(R)}$ does not depend on some particular combinations of fields.
On this subspace there is no constraint like \eq{eq:truncationconst}.

So after all it is only possible to get a truncation of the action, if
\eq{eq:truncationconst} is fulfilled on that subspace of $\phi$'s, on
which the highest term of the action, $s^{(R)}$, depends.  Keeping
translational invariance, it is impossible to have $s^{(R)}$
independent of fields at particular lattice points $n$, but $s^{(R)}$
may be independent of a whole field component ($\phi^i$) of the
multiplet.  Such a case appears in the case of constant fields (see
eq.\ \eq{eq:h_series}) when $a_1$ is set to zero.  Then the highest
term of the action, $\chi^4$, depends only on $\chi$, and
$M\alm+(M\alm )^T$ has no matrix entries for this field component.  In
this way a polynomial solution can be achieved.

We stress that so far we only discussed one necessary condition for a
truncation of the action down to a polynomial.  The remaining
equations \eq{eq:General0}-\eq{eq:General3} need to be solved as well.

The continuum limit of the resulting actions needs in general a
careful investigation since additional interaction terms must be
introduced to solve the relation. These terms have no corresponding
continuum counterparts and should vanish in the continuum limit.  It
is easy to ensure this in a naive way, where one just introduces the
appropriate power of the lattice spacing, $a$, in front of the terms
to let them vanish when $a$ goes to zero.  Notice, however, that the
additional interaction terms introduce new vertices in a perturbative
expansion.  Because of divergences (for $a\rightarrow 0$) the
corresponding diagrams are not necessarily vanishing even though the
vertices themselves vanish.  So it seems non-trivial to find the
correct form of the action in the continuum limit.

The situation gets even more difficult if the truncation constraint is
not fulfilled.  Then one has to accept a non-polynomial actions.
Beside the feasibility of such actions for numerical simulations, one
has to analyse fundamental aspects of them like renormalisability.
Nevertheless, these solutions obey the deformed lattice symmetry that
approaches the continuum one in the continuum limit.  This limits the
possible actions, e.g.\ by relating the different couplings, and might
even determine the entire form of the action.

How the discussion specialises to supersymmetry
will be discussed in the next subsection.

\subsection{Interacting SUSY}

To illustrate the above rather formal general statements, we specify
them to SUSY theories.  For the sake of the argument we consider now a
higher order action of SUSYQM, other SUSY theories will lead to a
similar situation.

We mimic the interaction term of the continuum theory,
$\bar{\psi}\psi\chi+F\chi^{2}$ (see
eq.~\eq{eq:action_cont} with $W=\chi^{2}/2$), by the following lattice
action:
\begin{eqnarray}
  s^{(3)}&=&
  K_{n_1,n_2,n_3}\bar{\psi}_{n_1}\psi_{n_2}\chi_{n_3}\nonumber\\
  &&+\,
  K'_{n_1,n_2,n_3}F_{n_1}
  \chi_{n_2}\chi_{n_3}
  \label{eq:action_higher_SUSYQM}
\end{eqnarray}
with parameters $K$ and $K'$ as in \eq{eq:actionseries}.

The (super)vector $v$ following from this action reads
\begin{equation}
 v^i_n=\left(
\begin{array}{c}
    K_{n_1,n_2,n}\bar{\psi}_{n_1}\psi_{n_2}+
  2 K'_{n_1,n_2,n} F_{n_1} \chi_{n_2}\\
K'_{n,n_1,n_2} F_{n_1} \chi_{n_2}\\
  -K_{n_1,n,n_2}\bar{\psi}_{n_1}\chi_{n_2}\\
  K_{n,n_1,n_2}\psi_{n_1}\chi_{n_2}
\end{array}\right)\, .
\end{equation}
It is not hard to see that $v$ takes all values in the considered
space of lattice fields upon varying the fields
$(\chi_n,F_n,\bar{\psi}_n,\psi_n)$ at all lattice sites.  The
rationale for this is, as explained above, the `completeness' of the
highest term in the action $s^{(3)}$,
eq.~\eq{eq:action_higher_SUSYQM}, if it contains all fields of the
multiplet. As a consequence, \eq{eq:truncationconst} has to be
fulfilled in general, but this reduces the symmetry relation
\eq{eq:SymWilson} to the naive symmetry.  One can easily convince
oneself that the same argument applies to all continuum-inspired
actions of the form ${(R-1)}\bar{\psi}\psi\chi^{R-2}+F\chi^{R-1}$.

There are two options to proceed at this point.  The first one is to
accept non-polynomial actions like \eq{eq:h_full}, which induces many
new complications.

The second and rather intricate possibility is to fulfil
\eq{eq:truncationconst} only on a subspace of the fields.  For SUSY
this can be achieved, e.~g., by building the highest order term (in
this case $s^{(3)}$) out of purely bosonic or purely fermionic fields.
As $\M\alm$ has only entries mixing bosons and fermions \eq{eq:vMv} is
fulfilled easily in this way.  We repeat the constant field result,
where setting $a_1=0$ and $a_0\neq 0$ in eq. \eq{eq:h_series} the
highest term is of the form $\chi^4$.  In the present case with
`smearing' in $K$, such a term could be represented as
$K_{n_1,n_2,n_3,n_4}\chi_{n_1}\chi_{n_2}\chi_{n_3}\chi_{n_4}$.

Of course, such incomplete parts of the actions on their own are not
giving the desired continuum limit since in the continuum SUSY actions
all fields are present at any given order (i.e.\ in every $s^{(r)}$).
The only meaning such terms could have is to be lattice artifacts
needed to solve the symmetry relation \eq{eq:SymWilson} at finite
lattice spacing and to vanish in the continuum limit, such that the
continuum action contains only terms of lower order in the fields.

Again, in this approach we so far have discussed only the highest
relation.  To solve the remaining relations is a non-trivial task to be
done in further investigations.

With the help of such additional terms it might be possible to fulfil
the symmetry relation which then guarantees the (super)symmetry in the
continuum limit.

\section{Summary}

In this work we have systematically approached lattice supersymmetry,
and have extended the Ginsparg-Wilson relation to general linear
symmetries and interacting theories.

For a general blocking procedure we have derived a lattice relation,
eq.~\eq{eq:SymWilson}, from the continuum symmetry of the theory at
hand. This relation can be viewed as the `remnant' of the continuum
symmetry on the lattice, and has to be satisfied by the lattice
action. The relation also includes potential anomalies and reduces to
the well-known Ginsparg-Wilson relation for chiral theories.

As this relation is derived for all possible blockings it does in
general not comprise a lattice symmetry, but rather describes the
breaking of the continuum symmetry by the blocking procedure. It can
be interpreted as a deformed symmetry only for those blockings for
which the symmetry operator in \eq{eq:ModSym} is local, as defined in
\eq{eq:genlocality}, and tends towards the continuum symmetry operator
in the continuum limit. These important properties are discussed in
section~\ref{sec:localityofMdef}. These requirements exclude for
instance the Wilson fermion action as a solution for a chiral lattice
theory.

Interestingly enough, the averaging function $f$ defines the lattice
formulation, but does not appear in the relation of the lattice
action.  Instead, $f$ is involved in an additional constraint
encountered in the derivation of the relation. This constraint is of
particular importance when the symmetry transformations include
derivatives, which is one of the characteristic features of
supersymmetry (as opposed to e.g.\ the algebraic
$\gamma_5$-transformations in the chiral case).
In order to keep the resolution of the averaging function as demanded
by the lattice cutoff, the derivative operator in the lattice
transformations must be the SLAC derivative or approach the SLAC
derivative in the continuum limit.  Such a non-local operator is
of course problematic in view of the locality of lattice action
and symmetry.

The other ingredient of the blocking transformation, the blocking
kernel $\alpha$, appears explicitly in the symmetry relation of the
lattice action.  An $\alpha$ that respects the symmetry leads to a
vanishing rhs of this relation and thus to a naive symmetry.  For
chiral theories this is forbidden because of vector symmetry.  In
case of supersymmetry a symmetric $\al$ can in principle be chosen,
but the resulting deformed symmetry would then contain the SLAC
derivative and hence be non-local.  The corresponding invariant lattice
actions inherit this problem.  Thus we are lead to non-symmetric
$\alpha$'s and a non-vanishing rhs of our relation to have a chance
to obtain local lattice actions and local symmetry transformations for
supersymmetry.

In the concrete example of quadratic SUSYQM one can find how $\alpha$
improves the locality of the action: An ultra-local $\alm$ introduces
propagator-like denominators in the lattice operators, which however
are not enough to compensate for the non-locality of the emerging SLAC
operator.  Considering $\alm$ as a momentum-dependent quantity allows
(ultra-) local lattice actions as a solution of the symmetry relation.
Beside the lattice action also the deformed lattice symmetry must be
local.  To get a connection to the blocking procedure, only a relaxed
version of locality can be guaranteed (cf.\ section
\ref{sec:local_symmetry}).  Hence this rather simple example already
reveals a lot about the interplay of the blocking ingredients, the
locality of actions and symmetries on the lattice.

One of the main differences between SUSY and chiral theories is that
the symmetry of the former acts on non-quadratic terms.
Correspondingly, the relation for lattice SUSY extends beyond second
order and couples different powers of fields.  We have discussed that
this generically results in non-polynomial actions.  This finding is
not completely surprising as our construction is half way towards the
full quantum effective action: we integrate out the original fields
and represent the remaining path integral in terms of the blocked
lattice fields. The non-polynomial actions are derived from the
continuum theory in a prescribed way.  One might for instance
speculate that certain cancellations between bosons and fermions are
still present and help to renormalise the system. We have given a
necessary condition for a truncation in the action.  Indeed, in the
very special example of constant fields in SUSYQM this criterion could
be satisfied and the action is local.

We conclude with a short outlook.  To date, the main problem for
performing SUSY lattice simulations is the implementation of the
symmetry.  We hope to use the blocking formalism to obtain local
interacting theories, which are invariant under a local deformed
symmetry.  The corresponding solutions to the symmetry relation should
either be obtained through reasonable approximations or by a
resummation.  This is work in progress.
\\[-1ex]

{\bf Acknowledgement}\\[-2.3ex]

\noindent We thank J.~Bloch, A.~Feo, C.~Gattringer,
H.~Gies, T.~K\"astner, G.~M\"unster, K.~Schoutens,
S.~Uhlmann, U.~Wenger, A.~Wipf, and in particular
Y.~Igarashi for discussions.  FB has been supported by DFG (BR
2872/4-1).  GB acknowledges support by the Evangelisches Studienwerk.

\hfill

\renewcommand{\thesection}{}
\renewcommand{\thesubsection}{\Alph{subsection}}
\renewcommand{\theequation}{\Alph{subsection}.\arabic{equation}}

\section*{Appendix} 
\setcounter{equation}{0}

\subsection{Solution for a quadratic action}
\label{sec:sfaqa}
The starting point to get the fixed point operator in its usual form
is the result of the Gaussian integration,
\begin{eqnarray}
K&=&\alpha-\alpha f (f^T\alpha f +\tilde{K})^{-1}f^T \alpha\nonumber \\
&=&\alpha\big[1-f (f^T\alpha f +\tilde{K})^{-1}f^T \alpha\big]\, .
\end{eqnarray}
A direct inversion of $f^T\al f+\tilde{K}$ is difficult since $f^T\al
f$ is, unlike $\tilde{K}$, non-diagonal in frequency space.  This
property comes from the $f$, that maps a larger space onto a smaller
one. So $f^T\al f$ acts on a larger space than the diagonal $\al$ and
is in general non-diagonal.  Therefore, one first gets a closed
expression of the inverse of $K$.  Assuming $\tilde{K}$ and $\alpha$
to be invertible an expansion in terms of a Neumann series yields
\begin{eqnarray}
  K^{-1}&=&\sum_{n=0}^\infty\big[f (f^T\alpha f +
  \tilde{K})^{-1}f^T \alpha\big]^n\alpha^{-1}\nonumber\\
  &=&\alpha^{-1}+f\sum_{n=0}^\infty\big[(f^T\alpha f +
  \tilde{K})^{-1}(f^T \alpha f+\tilde{K}-\tilde{K})\big]^n\nonumber\\
  &&\qquad\qquad\qquad\qquad\quad\times(f^T\alpha f +
  \tilde{K})^{-1}f^T\nonumber\\
  &=&\alpha^{-1}+f\sum_{n=0}^\infty\big[1-(f^T\alpha f +
  \tilde{K})^{-1}\tilde{K}\big]^n\nonumber\\
  &&\qquad\qquad\qquad\qquad\quad\times(f^T\alpha f +
  \tilde{K})^{-1}f^T\, .
\end{eqnarray}
After a resummation the result becomes
\begin{eqnarray}
  K^{-1}&=&\alpha^{-1}+f\big[(f^T\alpha f +\tilde{K})^{-1}
  \tilde{K}\big]^{-1}\nonumber\\
  &&\qquad\qquad\qquad\qquad\quad\times(f^T\alpha f +
  \tilde{K})^{-1}f^T\nonumber\\
  &=&\alpha^{-1}+f\tilde{K}^{-1}f^T \, .
\end{eqnarray}
An inversion of this expression is the know result mentioned in
equation~\eq{eq:BietenK}.

An other way to derive it is done via the introduction of auxiliary
fields $\sigma_n^i$,
\begin{eqnarray}
  e^{-\Sw[\phi]}&=&\int d\varphi\, e^{-\half(\phi-f\vp) 
    \alpha(\phi-f\vp)-\half\vp \tilde{K} \vp}\nonumber\\
  &=&\mathcal{N}^\pr\int d\varphi\, \int d\sigma\,
  e^{-\half\vp \tilde{K}\vp+i \sigma(\phi-f\vp)-
    \half\sigma\alpha^{-1}\sigma}\nonumber\, .
\end{eqnarray}
After a Gau\ss ian integration (first of the $\vp$ then of the
$\sigma$) one gets again the desired expression for $K$
\begin{eqnarray}
  e^{-\Sw[\phi]}&=&\mathcal{N}^{\pr\pr} e^{-
    \half\phi(f\tilde{K}^{-1}f^T+\alpha^{-1})^{-1}\phi}\nonumber\\
  &=&\mathcal{N}^{\pr\pr} e^{-\half\phi_n^i K_{nm}^{ij}\phi_m}.
\end{eqnarray}
With the same conventions as in section \ref{sec:FTconst} one can
easily derive the Fourier representation of $K$.

\subsection{Fourier analysis of the additional constraint}
\label{sec:FTconst}

For simplicity we work in a one-dimensional finite volume of size $L$.
A general continuum field has the Fourier series representation
\begin{eqnarray}
 \varphi(x)&=&\sum_{q=-\infty}^\infty\varphi(p_q)e^{i p_q x}\\
 \varphi(p_q)&=&\frac{1}{L}\,\int_0^L \!dx\, \varphi(x)e^{-i p_q x}\, ,
\end{eqnarray}
with dimensionless wave numbers $q$ ($\in \mathbb{Z}$) and $p_q=\frac
{2 \pi q}{L}$.  The same representation can be applied to the
averaging function $f(an-x)$.

We discretise $L=Na$ with an odd number $N$ of lattice points and
lattice spacing $a$.  Functions on the lattice can be parametrised by
$N$ independent waves which we choose to be in the first Brillouin
zone,
\begin{eqnarray}
 \phi(an)&=&\sum_{k=-(N-1)/2}^{(N-1)/2}\phi(p_k)e^{i p_k a n}\\
 \phi(p_k)&=&\frac{1}{N}\sum_{n=0}^{(N-1)}\phi(an)e^{-i p_k an}\, .
\end{eqnarray}
From this relation it is clear that $\phi(p_k)$ is periodic in $p_k$,
$\phi(p_k)=\phi(p_k+l 2\pi/a)=\phi(p_{k+lN})$ $\forall l\in
\mathbb{Z}$. The same transformation is used for the $\phi_{f}(an)$.
(For $\phi_f(p_k)$ $k$ runs from $-(N-1)/2$ to $(N-1)/2$.)

In Fourier space the convolution in the averaged field of
eq.~\eq{eq:average} becomes a product,
\begin{eqnarray}
  \phi_f(p_k)&=&\frac{L}{N}\sum_{n=-\infty}^{\infty}
  \sum_{m=0}^{(N-1)} e^{i(p_q-p_k)an}f(p_q)\vp(p_q)\nonumber\\
  &=&L\sum_{l=-\infty}^\infty f(p_k+l \frac{2\pi}{a})
  \varphi(p_k+l\frac{2\pi}{a})\,.
\label{eq:fourier_product}
\end{eqnarray}
This shows how the averaging projects the Fourier components of $\vp$
onto the first Brillouin zone. In addition one easily observes that
the Fourier components of $\phi_f$ and the lattice fields are
determined by $f$, which means that $f$ introduces a cutoff for the
lattice momentum if $f(p_q)$ vanishes for all $p_q$ greater than the
cutoff.

Because of the constraint \eq{eq:constraint_final_trivial}, 
the sum on the rhs actually has at most one term.

The additional constraint \eq{eq:intconst} reads after partial
integration
\begin{equation}
 \sum_m \n_{nm}f(am-x)+\partial_x f(an-x)=0\quad \forall\, n,x\,.
\end{equation}
Because of the circulant form of $\n$ (cf.\ \Eq{eq:circulant}) 
we define its Fourier transform as
\begin{equation}
  \n_{nm}=\frac{1}{N}\sum_{k=-(N-1)/2}^{(N-1)/2}\n(p_k)\,
  e^{i p_k a(n-m)}\,,
\end{equation}
with
\begin{equation}
  \n(p_k)\delta_{kl}=\frac{1}{N}\sum_{m=0,\, n=0}^{(N-1)}\n_{nm}
  e^{i p_k an}e^{i p_l am}\, ,
 \label{eq:nabla_on_wave}
\end{equation}
where $k$ and $l$ are integer numbers between $-(N-1)/2$ and
$(N-1)/2$.  For greater values of $k$ the result of the Fourier
transformation again fulfils $\n(p_{k+lN})=\n(p_k)$ and
$\delta_{kl}$ has the same periodicity (which means it is one only for
$k=l \mod N$ and otherwise zero).

With these Fourier transforms the constraint becomes
\begin{equation}
  \sum_{q=-\infty}^{\infty}f(p_q)\Big[ \n(p_q)-i p_q\Big] 
  e^{i p_q (an-x)}=0\, ,
\end{equation}
which for every individual component $p_q$ gives the constraint
\eq{eq:constraint_final}.

\subsection{Doubler problem}
\label{sec:app_doublers}

In this appendix we discuss solutions $\n$ of the additional
constraint which are ultra-local.  These include the (slightly
modified) symmetric difference, \eq{eq:n_symm}, \eq{eq:c_first} and
extensions thereof spreading over neighbours at larger distances, e.g.\
\eq{eq:next_best1}.  They come with blocking functions $f$ extended
over the whole volume, \eq{eqn_f_symm}, or subvolumes, depending on
how many Fourier components $f$ can have.  In Section~\ref{sec:cont_f}
we have argued that only the $f$ with the highest resolution $O(a)$
leads to a reasonable, namely non-redundant lattice theory.  This
amounts to $\n$ being the SLAC operator.

Although the use of these ultra-local $\n$'s is questionable,
we nevertheless want to point out that doublers in them can be removed.
For supersymmetric theories it is natural that also bosons have a
doubling problem \cite{Catterall:2000rv},
so we investigate the kinetic operator in the bosonic sector, too.

Consider the solutions \eq{eq:soln_tilde_nabla}-\eq{eq:soln_boxmb}.
The operator $\n$ in them has doublers at the edge of the Brillouin zone, 
unless $\n$ is the SLAC-operator, 
and $\tn$ inherits them via the solution \eq{eq:soln_tilde_nabla}.

To resolve the doubler problem we make use of the fact, that
the matrix $m_b$ in our ansatz \eq{eq:action_quadratic_SUSYQM} 
and in \eq{eq:soln_tilde_nabla}-\eq{eq:soln_boxmb}
is only restricted to be symmetric and circulant.  
In the spirit of the Wilson term, we will now replace $m_b$ by
\begin{equation}
  (m_b)_{nm}=m\delta_{nm}+\frac{r}{2 a}(2\delta_{nm}-
  \delta_{n+1,m}-\delta_{n-1,m}) 
\end{equation}
with $r$ a dimensionless parameter.
Such a momentum-dependent correction will then also occur in $m_f$ 
through our solution\footnote{ 
  In our setting, where $m_f(m_b)$, it is the
  simplest to change $m_b$.  One could also have solved $m_b(m_f)$ and
  then change $m_f$.}  
and all -- bosonic and fermionic -- doublers
are removed as long as the parameter $a_2$ remains small.

The dispersion relations of $m_b$ and $\n$ are
\begin{eqnarray}
 m_b(p_k)&=&m+\frac{r}{a}(1-\cos(ap_k))
 \label{eq:dispersion_relation1}\\
 \n(p_k)&=&\frac{i}{a}\sum_{l=1}^{(N-1)/2}c_l\,\sin(l a p_k)\,.
 \label{eq:dispersion_relation2}
\end{eqnarray}

\begin{figure*}
 \includegraphics[width=16cm]{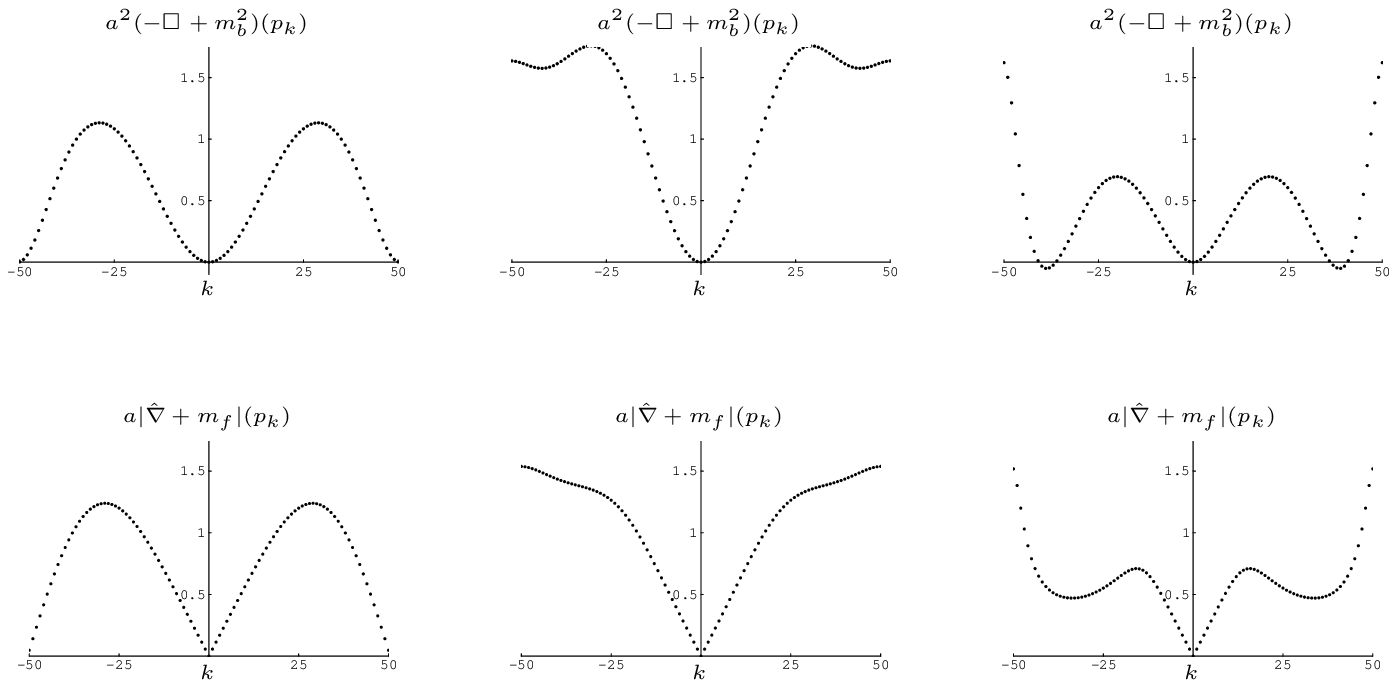}
 \caption{Dispersion relations for the bosonic and fermionic bilinears
   with ultra-local $\n$ and Wilson mass in $m_b$, cf.\ equations
   \eq{eq:dispersion_relation1} and \eq{eq:dispersion_relation2}, on
   an $N=101$ lattice.  In the top row we plot the bosonic
   $-\Box+m_b^2$ and in the bottom row the fermionic $|\tn+m_f|$,
   according to the solutions
   \eq{eq:soln_tilde_nabla}-\eq{eq:soln_boxmb}, as a function of the
   momentum $p$ in the first Brillouin zone.  On the left: $r=0$ with
   doublers visible at the edges.  In the middle: the doublers are
   removed by $r=1$ and $(a_0,a_1,a_2) =(0.1,0.1,0.3)$.  On the right:
   additional zeroes in the bosonic sector occur for the choice $r=1$
   and $(a_0,a_1,a_2)=(0.1,0.1,2.0)$.}
 \label{fig:disp_relations}
\end{figure*}

We consider massless fields $m=0$.
Then because of $m_b(0)=0$ and $\n(0)=0$
it is clear that the bilinears 
\begin{equation}
 \frac{1}{2}\,\chi(-\Box+m_b^2)\chi\,,\quad\bar{\psi}(\tn+m_f)\psi
\end{equation}
according to \eq{eq:soln_tilde_nabla}-\eq{eq:soln_boxmb} have zeroes
at vanishing $p$ as they should.  At the edge of the Brillouin zone,
$p=\pi/a$, they would vanish as well because of $\n(\pi/a)=0$, if
there were no $r$-corrections.  The finite value $m_b(\pi/a)=2r/a$,
however, removes the doublers since
\begin{eqnarray}
  (-\Box+m_b^2)(p=\frac{\pi}{a})&=&\left(\frac{2r}{a}\right)^2
  \frac{1}{1+a_0}\\
  (\tn+m_f)(p=\frac{\pi}{a})&=&\frac{2r}{a}\frac{1}{1+2ra_1/a+a_0}
\end{eqnarray}
These values are non-zero and scale depending on the behaviour of the
$a_i$.

What remains to be shown is that there are no other zeroes at
intermediate values of $p$.  For that we specialise to the solution of
$\n$ ranging to next-to-nearest neighbours with $c_{1,2}$ from eqn.s
\eq{eq:next_best_c1},\eq{eq:next_best_c2} and plot in Fig.\
\ref{fig:disp_relations} the operators $-\Box+m_b^2$ and $\tn+m_f$ in
momentum space.  As $-\Box+m_b^2$ and $m_f$ are real symmetric (also
after including $r$), their Fourier transforms are real, whereas that
of the real antisymmetric $\tn$ is purely imaginary.  In order to seek
for zeroes, we therefore plot $(-\Box+m_b^2)(p)$ and
$|(\tn+m_f)(p)|=\sqrt{\tn^2(p)+m_f^2(p)}$, both as functions of $p$.
As one can see from that figure, the doublers are removed and no
additional zeroes appear, unless $a_2$ is too big.

\subsection{Alternative derivation of the symmetry relation}
\label{sec:without_alpha}

Let us discuss how the symmetry relation emerges in the setting of
section \ref{sec:contlimgen} with currents.  Here we take the point of
view that the Wilsonian action $S$ is defined to give the same
generating functional $Z_f[J]$ in \eq{eq:fullZ},
\begin{eqnarray}
\label{eq:fullZ_recap} 
Z_f[J] = \0{1}{\Norm e^{\s012 J\alpha^{-1} J}} 
\int\d\phi\, e^{-\Sw[\phi]+ J\phi }\,,
\end{eqnarray} 
for all currents $J$  as the classical action in \eq{eq:genfuncblock},
.\begin{eqnarray}
\label{eq:genfuncblock_recap} 
Z_f[J]=\0{1}\Norm\int
d\varphi \, e^{-\Sc[\varphi]+ J \phi_f[\varphi]} \,. 
\end{eqnarray}
This yields the same expectation values for all observables
composed out of the blocked fields $\phi_f$.

Now performing a symmetry variation
$\varphi\to\varphi+\epsilon\Mc\varphi$ on the classical fields
$\varphi$ and changing the integration variable in
\eq{eq:genfuncblock_recap} accordingly, one gets to $O(\epsilon)$:
\begin{eqnarray}
\!\!\!\!\!\!\!\epsilon\Str\Mc Z_f[J]=\0{1}{\Norm}\int
d\varphi \, e^{-\Sc[\varphi]+ J \phi_f[\varphi]}\epsilon J \phi_f[\Mc\varphi]
\end{eqnarray}
The second term can be rewritten as $\epsilon J\M\phi_f[\varphi]$ 
using the additional constraint.
Then it is just the change of $Z_f[J]$ under 
$J\to J+\epsilon J M$,
\begin{eqnarray}
 \epsilon\Str\Mc Z_f[J]=Z_f[J+\epsilon J M]-Z_f[J]
\end{eqnarray}
This can be computed in the representation \eq{eq:fullZ_recap}, where
again we change the integration variable $\phi\to\phi+\epsilon M
\phi$.  This yields a $\Str\M$ part, the variation of $S$ and a
bilinear term in $J$ from the variation of the prefactor:
\begin{eqnarray}
 \!\!\!\!\!0=\epsilon\Big\{\Str\Mc - \Str\M+\delta S - JM\alm J\Big\} Z_f[J]
\end{eqnarray}
The currents in the last term can be rewritten as $\phi$-derivatives
of $\exp(J\phi)$ in $Z_f[J]$ and, by partial integration, of
$\exp(-S[\phi])$:
\begin{eqnarray}
  &&0=\epsilon\int\d\phi\, e^{-\Sw[\phi]+ J\phi }
  \Big\{\Str\Mc - \Str\M+\delta S\nonumber\\
  &&- e^{\Sw[\phi]}\frac{\partial}{\partial\phi}M\alm 
  \frac{\partial}{\partial\phi}e^{-\Sw[\phi]}\Big\} Z_f[J]
\end{eqnarray}
Demanding this equality for all currents $J$, the curly bracket in the
integrand has to vanish giving \eq{eq:deltaWilson1} and thus our
symmetry relation \eq{eq:SymWilson}.

Notice that we have used the same ingredients as in the original
derivation of the relation, including the additional constraint and
the inverse blocking kernel $\alm$, but we have not relied on the
definition \eq{eq:SWilsonrecap} of the Wilsonian action based on
$\alpha$ itself.

\subsection{Singular $\alm$}
\label{app:singalm}
The results of section~\ref{sec:contlimgen} extend to $\alpha$'s with
some diverging matrix elements $\alpha_{mn}$. The corresponding
eigenmodes $\alpha \phi_{\rm sing}=\lambda_{\rm sing} \phi_{\rm sing}$
with $\lambda_{\rm sing}\to\infty$ are fixed: $\phi_{\rm
  sing}=\phi_{\rm sing,f}[\varphi]$. The remaining modes in $\phi$
still undergo the smearing procedure.

The other extremal case is $P_{\rm sing}{\alpha^{-1}} P_{\rm sing}\to
\infty$, where $P_{\rm sing}$ is the projection operator on the
singular part of $\alpha^{-1}$. Still \eq{eq:alpha} is valid but the
definitions of the Wilsonian action \eq{eq:SWilsonrecap} and the
normalisation \eq{eq:norm} develop singularities. If we want to keep
these definition we could simply change the generating functional
$Z_f[J]$ to $Z_f[(1-P_{\rm sing})J]$ which generates correlation
functions $\langle \phi_{f,\rm reg}\cdots \phi_{f,\rm reg}\rangle$.
Here $\phi_{f,\rm reg}= P_{\rm reg}\phi_f$ with $P_{\rm reg}=(1-P_{\rm
  sing})$. Then the above derivations are unaltered with $\alpha\to
\alpha_{\rm reg}=P_{\rm reg}\alpha P_{\rm reg}$ and $\alpha^{-1}_{\rm
  reg} = P_{\rm reg}\alpha^{-1} P_{\rm reg}$. With this procedure we
have removed the $\phi_{\rm sing}$-modes from our theory. In
particular the definition of the Wilsonian action \eq{eq:SWilsonrecap}
with $\alpha_{\rm reg}$ leads to $S[\phi]=S[\phi_{\rm reg}]$.


\subsection{General blocking matrices $\alpha_G$}
\label{sec:GenAlpha}

The restrictions of the blocking matrix originate from the hermiticity
of the lattice action.  In addition, the most general blocking matrix
$\alpha_G$ should connect fermionic and bosonic fields only among each
other.  Therefore, the form of its inverse is restricted to be
\begin{multline}
  a\,(\alm_G)^{ij}_{mn}=\\
  \left(\begin{array}{c c c c}
      \elm_2&\elm'_1&0&0\\
      \elm'_1&\elm_0&0&0\\
      0&0&0&\sy{\elm_1}+\asy{\elm_1}\\
      0&0&-\sy{\elm_1}+\asy{\elm_1}&0
\end{array}\right)_{mn}\, ,
\label{eq:alpha_minus_one_SUSY}
\end{multline} 
where $\elm_1''$ has to be antisymmetric, whereas
all other matrices are symmetric.  In order to get translation
invariance all matrices must be circulant.

The most general symmetric matrix, fulfilling \eq{eq:vanish} with the symmetry operators $M$ and $\bar{M}$ defined
by \eq{eq:M} and \eq{eq:Mbar}, is 
\begin{multline}
\label{eq:MalmS}
a(\alm_S)^{ij}_{mn}=\\
\left(
\begin{array}{cccc}
  \n^{-1}\asy{\elms} & \sy{\elms} & 0 & 0 \\
  \sy{\elms} & \n \asy{\elms} & 0 & 0 \\
  0 & 0 & 0 & -\sy{\elms}+\asy{\elms} \\
  0 & 0 & \sy{\elms}+\asy{\elms} & 0
\end{array}
\right)_{mn},
\end{multline}
where $\asy{\elms}$ is an antisymmetric and $\sy{\elms}$ a symmetric
circulant matrix.  Such a symmetric part, $\alm_S$, can always be added to
$\alm$ without changing the relation \eq{eq:quadGW}.  So the matrix elements of
\begin{equation} 
\alm=\alm_G+\alm_S
\end{equation} 
can be reduced without any loss of generality concerning the modified
symmetry using the freedom of choosing $\asy{\elms}$ and $\sy{\elms}$.
We choose them to be
\begin{equation}
 \sy{\elms}=-\elm_1\quad \text{and}\quad \asy{\elms}=-\asy{\elm}_1
\end{equation}
and define
\begin{equation}
  a_0:=\elm_0-\n\asy{\elm}_1\,, \; a_1:=\sy{\elm}_1+
  \elm_1\, , \; a_2:=\elm_2-\n^{-1}\asy{\elm}_1\, , 
\end{equation}
where all $a_i$ are now symmetric circulant matrices.  The resulting
matrix with reduced matrix elements compared to $\alm_G$ is just the one given in equation \eq{eq:redAlm}.
Since $\alm_S$ is not important for the deformed symmetry of the action the result is not changed compared to $\alm_G$.
However, the deformed symmetry $M_{\rm def}$ depends on this part. So if one wants to investigate the locality of $M_{\rm def}$ the most general blocking matrix $\alm_G$ instead of the $\alm$ must be considered. 

From these considerations it is clear that all entries of $\alm$ can in principle be zero after
splitting off $\alm_S$.  Then the blocking matrix would consist only
of a symmetric part $\al_S$.  This was done in \cite{So:1998ya} where
$\alm=\alm_S $ with $\asy{\elms}=0$ was used.  The lattice symmetry is
in that case reduced to the naive one.

\subsection{Decay of SUSY lattice operators}
\label{app:propagator}

In this appendix we discuss the locality properties of the operators
\eq{eq:soln_tilde_nabla}-\eq{eq:soln_boxmb} with momentum-independent $a_i$,
which due to the denominators resemble massive propagators. 
Nonetheless, we will demonstrate that the position space representations 
of these operators do not decay exponentially on the lattice.

Let us, as common in the discussion of locality, consider the limit of
large lattices.  That is, we replace $\n$ in eqn.\
\eq{eq:soln_tilde_nabla} by $p_k$ for $k=-(N-1)/2\ldots (N-1)/2$ and
take $N$ to infinity.  Then for the position space representation of
the (circulant) fermionic operator $\tn$
\begin{equation} 
\tn_{nm}=\frac{1}{2a}\sum_{l=-(N-1)/2}^{(N-1)/2}
\hat{c}_{l}\,\delta_{n-m ,-l}\, ,
\end{equation} 
cf. \eq{eq:circulant},
the discrete Fourier transformation turns into a Fourier integral
\begin{eqnarray}
  \hat{c}_l&=&\lim_{N\to\infty}\frac{1}{N}\!\!\!\sum_{k=
    -(N-1)/2}^{(N-1)/2}\!\!\! \tn(p_k)
  e^{i p_k l}\nonumber\\
  &=&\int_{-\pi/a}^{\pi/a}\!\!\!dp\, \tn(p)\,e^{ipla}\, ,
 \label{eq:fourier_locality}
\end{eqnarray}
where $p$ is a continuous momentum and
\begin{equation}
  \tn(p)=i\frac{(1+a_0-a_2m_b^2)p}{(1+a_1m_b+a_0)^2+(a_2m_b+a_1)^2 p^2}\, .
\end{equation}
from the solution \eq{eq:soln_tilde_nabla}.

In the limit $a\to 0$ and with $p$-independent $a_i$, 
the Fourier transform gives an exponential decay because
\begin{eqnarray}
\!\!\int_{-\infty}^\infty\!\!\!dp \frac{p}{\eta_1^2+\eta_2^2 p^2}e^{ipla}\sim
 \exp\left(-{\rm sign}(l)\,\frac{la|\eta_1|}{|\eta_2|}\right).
\end{eqnarray}

At finite lattices the Fourier transform \eq{eq:fourier_locality} 
can be computed in the complexified $p$-space.  
The Fourier integral consists of contributions from poles where the
denominator of $\tn(p)$ vanishes and an additional part from the
corresponding complex contours that encloses the poles.

Because of the regularity of $\tn(p)$ on the real axis, the poles
appear away from it.  For positive $l$ we enclose the poles with
positive imaginary parts by adding paths $p=\pm\pi/a +
i\zeta,\,\zeta\in[0,\infty)$.  The residue at each pole is some number
coming from $\tn$, which we assume to be non-zero, times a term
decaying exponentially in $l$ from the Fourier factor in
\eq{eq:fourier_locality}.  For the same reason there is no
contribution from the asymptotic contour at large Im $p$ parallel to
the real interval $[-\pi/a,\pi/a]$.

The two new contours contribute
\begin{equation}
\pm\int_0^\infty \!d\zeta\, \tn(p)\,e^{ipla}\Big|_{p=\pm\pi/a+i\zeta}\, ,
\end{equation}
which turn into Laplace transforms
\begin{equation}
  \int_0^\infty\!d\zeta\, \big[\tn(\pi/a+i\zeta)-\tn(-\pi/a+i\zeta)
  \big]e^{-\zeta l a} \, .
\end{equation}
For large spatial distances $l$ only the value of $\tn$ at $p=\pi/a$
remains, namely
\begin{equation}
\big[\tn(\pi/a)-\tn(-\pi/a)\big]\int_0^{1/la}\!\!\!d\zeta (1-\zeta l a)
=\tn(\pi/a)\,\frac{1}{la}
\end{equation}
where we used that $\tn$ is odd.  Hence the additional contours give
\emph{algebraic corrections to $\tn$ in position space}, unless $\tn$
vanishes at the boundary of the Brillouin zone $p=\pi/a$.  Of course
the problem of algebraic tails is absent, if $\n$ is ultra-local.
Here, however, we have considered $\tn$ as a function of the SLAC
operator $\n$ according to the solution \eq{eq:soln_tilde_nabla}.  The
way out, which is followed in the body of the paper, section
\ref{sec:locality}, is to consider general momentum-dependent blocking
kernels.

We remark that the discussion is similar for negative $l$ (where one
encloses the poles with negative imaginary parts).  Furthermore, the
algebraic corrections do not appear in $m_f$ and $-\Box+m_b^2$, eqn.s
\eq{eq:soln_mf},\eq{eq:soln_boxmb}, because these are even functions in
$p$.

\subsection{Locality condition}
\label{app:local}
All of the following one dimensional considerations can be extended
easily to higher dimensions.  The considered lattice operators $O$
are, because of translational invariance, circulant matrices,
\begin{equation}
 O_{mn}=O_{n-m}=F(a(n-m))\, .
\end{equation}
The slightly modified condition for locality demands 
that $F$ decays faster than any polynomial. 
That means
\begin{equation}
  |x^r F(x)| < \infty\quad \forall\, r \in 
  \mathbbm{N},\, x,y \in a\mathbbm{N}\, .
\end{equation} 
If the Fourier transform of $F(x)$, $f(p)$, and its derivatives have
no singularities and fulfil periodic boundary conditions at edge of
the BZ the following estimation can be made
\begin{eqnarray}
 |x^r F(x)|&=&|\int_{\rm BZ} (\pa_p^r f(p))e^{ipx}|\nonumber\\
&\leq& \int_{\rm BZ} |\pa_p^r f(p)|\leq C_r<\infty\, .
\end{eqnarray}
Consider now a non-local operator similar to the SLAC derivative.
This non-local operator should have no singularities within the BZ for
all of its derivatives.  The boundary conditions are, however, not
periodic.  According to the discussion of the locality of $K$ and
$M_{\rm def}$ it should support the modified locality after a
multiplication with a local operator.  In view of the above argument
the boundary conditions must hence be enforced by this local operator
without spoiling the differentiability of $f$.  Its representation in
Fourier space, $I(p)$, must therefore be a function that vanishes
together with all its derivatives at the edge of the BZ. In addition
no singularities should appear within the BZ for any of its
derivatives. One function that fulfils these requirements is
\begin{equation}
  I(p)=\begin{cases}
    \exp\left(-\frac{\epsilon^2}{\epsilon^2-p^2}\right) &
    |p|<\epsilon\\0&|p|\geq\epsilon
\end{cases}\; \mathrm{with}\; \epsilon\leq\frac{\pi}{a} .
\end{equation}
It is clear that $I(p)$ can not be analytic since any analytic
function that vanishes with all its derivatives at a specific point
must be identical to zero. So the common definition of locality in
terms of analyticity in momentum space can not be satisfied.


\end{document}